\documentclass[12pt]{article}
\usepackage[left=2.5cm,top=2.5cm, bottom=1.2in, right=2.5cm]{geometry}
\usepackage[english]{babel}
\usepackage[cmex10]{amsmath}
\usepackage{amssymb, mathrsfs, amsfonts, amsthm}
\usepackage{mathtools}

\usepackage[usenames,dvipsnames]{xcolor}
\usepackage{graphicx}
\usepackage{paralist}

\usepackage[colorlinks=true]{hyperref}
\usepackage{textcomp}
\usepackage{caption}
\usepackage{subcaption}
\usepackage{stmaryrd}

\usepackage{float}
\usepackage{grffile}

\usepackage{algorithm}
\usepackage{algpseudocode}
\makeatletter 
 
\@addtoreset{algorithm}{chapter} 
\makeatother

\usepackage{pgfplots}
\usepackage{tikz}
\usepackage{float}
\usetikzlibrary{plotmarks}
\usetikzlibrary{arrows.meta}
\usetikzlibrary{positioning}
\usepgfplotslibrary{patchplots}
\pgfplotsset{compat=newest}

\newtheorem{theorem}{Theorem}
\newtheorem*{theorem*}{Theorem}
\newtheorem{lemma}{Lemma}
\newtheorem*{lemma*}{Lemma}

\newtheorem*{proposition*}{Proposition}
\newtheorem{corollary}{Corollary}
\newtheorem*{corollary*}{Corollary}
\newtheorem{definition}{Definition}
\newtheorem*{definition*}{Definition}

\newtheorem*{example*}{Example}
\newtheorem{remark}{Remark}
\newtheorem*{remark*}{Remark}

\renewcommand{\H}[1]{H\!\left( #1 \right)}
\newcommand{\Hb}[1]{H\big( #1 \big)}

\newcommand{\I}[2]{I\!\left( #1 ; #2 \right)}

\newcommand{\E}[1]{\mathbb{E}\left[{#1}\right]}

\def\indep{\:\text{\mbox{$\perp\!\!\!\perp$}}\:}
\newcommand{\floor}[1]{\left\lfloor #1 \right\rfloor}

\def\pmf{\mathop{\mathrm{PMF}}}
\def\si{\mathop{\mathrm{SI}}}
\def\rv{\mathop{\mathrm{RV}}}

\allowdisplaybreaks
\allowbreak

\usepackage{IEEEtrantools}
\usepackage[toc,page]{appendix}

\begin{document}
\title{Secure Source Coding with Side-information at Decoder and Shared Key at Encoder and Decoder}
\author{Hamid Ghourchian, 
	Photios A. Stavrou, 
	Tobias J. Oechtering, \\
	and Mikael Skoglund 
	\thanks{This is a long version of the accepted ITW 2021 paper. This work was supported in part by the Swedish Strategic Research Foundation.}
	\thanks{The authors are with the Department of Intelligent Systems, Division of Information Science and Engineering, at KTH Royal Institute of Technology, 10044 Stockholm, Sweden (e-mail: \mbox{\{hamidgh; fstavrou; oech; skoglund\}@kth.se}).}
}

\renewcommand\footnotemark{}
\renewcommand\footnoterule{}

\date{}
\maketitle

\begin{abstract}
We study the problem of rate-distortion-equivocation with side-information only available at the decoder when an independent private random key is shared between the sender and the receiver.
The sender compresses the sequence, and the receiver reconstructs it such that the average distortion between the source and the output is limited.
The equivocation is measured at an eavesdropper that intercepts the source encoded message, utilizing side-information correlated with the source and the side-information at the decoder.
We have derived the entire achievable rate-distortion-equivocation region for this problem.
\newline\newline
\textbf{Keywords---}
	Source coding, rate-distortion, security, side-information, shared key.
\end{abstract}

\section{Introduction} \label{sec:intro}
\makeatletter
\if@twocolumn
\IEEEPARstart{I}{n}
\else
In
\fi
\makeatletter
 this paper, we consider a source (sequence of random variables ($\rv$s)) that is given to the sender.
The receiver does not have access the source, but it has access to a correlated sequence of $\rv$s that serves as side-information ($\si$).
In addition, both the sender and the receiver have access to a secure sequence of random bits (shared-key).
The encoder at the sender, compresses the source using the shared-key such that the legitimate receiver is able to estimate the source using the $\si$ and the shared-key, with some limited distortion.
In our setup, there exists a passive eavesdropper who has access to the output of the encoder and a possibly different $\si$, as depicted in Fig. \ref{fig:system}.
The goal is to minimize the output rate of the encoder while the distortion between the source and its estimated value, as well as the leaked information from the source to eavesdropper, are limited.
This setup is a special case of a more general setup when all the sender, receiver, and the eavesdropper have different versions of $\si$.
Here, the shared-key can be considered as the common $\si$ between sender and receiver that is not available to the eavesdropper.

\makeatletter
\if@twocolumn
\else
\subsection{Literature Review}
\fi
\makeatother
\makeatletter
\if@twocolumn
\else
Classical rate-distortion theory was introduced by Shannon in \cite{Shannon59}.
It identified the trade-off between the minimum achievable distortion and the rate of non-causal encoder and decoder pair. 
For a complete overview, one can see, for instance, \cite{berger:1971}.
\fi
\makeatother

Shannon introduced the notion of communication secrecy from an information-theoretical perspective in \cite{Shannon49}.
During the 70s, Wyner introduced the wiretap channel \cite{Wyner75} and showed that it is possible to send information at a positive rate with perfect secrecy when eavesdropper's channel is a degraded version of the channel from the encoder to the decoder.
When it comes to secrecy using information-theoretic tools, often two approaches can be found in the literature.
The first one presupposes that both encoder and decoder agree on a secret key before the transmission.
The second one assumes that the decoder and the eavesdropper (sometimes the encoder as well) have different versions of $\si$, and thereby secrecy is achieved through this difference.
For instance, Shannon in \cite{Shannon49} adopted the first approach and showed that the transmission of a discrete memoryless source is entirely secure if the rate of the key is at least as large as the entropy of the source.
Yamamoto in \cite{Yamamoto97} studied various secure source coding scenarios that include, among other results, an extension of Shannon's cipher system to combine secrecy with rate-distortion theory.

Prabhakaran and Ramchandran in \cite{Prabhakaran07} considered lossless source coding with $\si$ at both the decoder and the eavesdropper when there is no rate constraint between the encoder and the decoder.
Gunduz et al. in \cite{Gunduz08}, the authors considered a setup with $\si$ at the encoder and coded $\si$ at the decoder.
Villard and Piantanida in \cite{Villard13} studied the problem of secure lossy source coding when one or both the receiver and the eavesdropper have $\si$ (Fig. \ref{fig:gensystem} with $W=\emptyset$).
Chia and Kittichokechai in \cite{Chia13} adopt a setup where there exists a common $\si$ at the sender and the receiver, while the $\si$ of the eavesdropper can be different (Fig. \ref{fig:gensystem} with $W=Y$).
They showed that under certain Markov chain assumptions, or log-loss distortion, the rate-distortion-equivocation region has a closed-form solution.
In \cite{Merhav08}, the author studied a joint source-channel coding problem where there is $\si$ at the decoder and the eavesdropper. 
The paper characterizes the solution when the $\si$ at the eavesdropper is a degraded version of the output of the channel in addition to a degraded version of the SI at the decoder.

In a different direction where delay constraints may appear in the system, Kaspi and Merhav in \cite{Kaspi15} considered two source coding models combining causal or zero-delay source coding under secrecy constraints.
The causal source coding was defined in \cite{Neuhoff82}.
Ghourchian et al. in \cite{Ghourchian21} considered secure lossy and lossless compression with sequential encoding and non-sequential decoding, such that the rates and equivocations are defined cumulatively over sequential blocks. 
They characterized the achievable rate profile-distortion-equivocation profile region.

There exist different approaches to characterize and obtain the leakage in source coding problems. 
For instance, Song et al. in \cite{Song14} defined a distortion-based equivocation; for lossless case, when the receiver's $\si$ is more capable, the rate-equivocation region is found.
Schieler and Cuff in \cite{Schieler14} studied the lossy case of \cite{Song14} under the assumption that both the transmitter and the receiver share a secret key, but the receiver does not have access to the $\si$ (Fig. \ref{fig:gensystem} with $W^n=Y^n=K$ which is a secret independent key and different $L$ which is a kind of additive distortion).
Kittichokechai et al. in \cite{Kittichokechai14}, studied the leakage from the output of the decoder to the eavesdropper instead of the leakage from the source to the eavesdropper (end-user privacy).
\makeatletter
\if@twocolumn
For some cases, the achievable region has been identified.
\else
$\si$ exists at the decoder and the eavesdropper. In some cases, such as the one that the estimation of the input should only be memoryless with respect to the $\si$ and the message, the complete region is identified.
\fi
\makeatother

\makeatletter
\if@twocolumn
Interesting directions of secure source coding with a helper can be found in \cite{Kittichokechai16}, \cite{Bross16}, \cite{Benammar16ITW}, and \cite{Lu21} and for multi-user setups in \cite{Tandon13}, \cite{Balmahoon15}, and \cite{Naghibi15}. 
\else
Secure source coding with multiple nodes has also been studied in recent years.
The authors of \cite{Tandon13} have studied the problem in which there exists a sender, two distributed receivers, one of which has a $\si$.
The goal is to maximize the equivocation of the $\si$ given the message sent to the receivers.
In \cite{Balmahoon15}, there are two distributed but correlated sources, each of which should be estimated, with limited distortion, at the same decoder, while the eavesdropper also has access to a $\si$.
The equivocation region of the two sources is desired.
In the CEO problem, two different noisy observations of the same source are observed by distributed agents. They compress their observations such that the decoder is able to estimate the source \cite[Section 12.4]{ElGamal11}.
The CEO problem with secrecy constraints has been studied in \cite{Naghibi15}, such that the eavesdropper knows the description of one of the agents.

Secure source coding with a helper node has also been studied.
The helper receives the description from another node, usually the encoder. 
Then compresses the description, possibly with the help of $\si$, and then sends the output to the decoder. 
Since there is also a direct link, it is called \emph{triangular helper} \cite{Kittichokechai16}.
In \cite{Kittichokechai16}, all sender, receiver, and the helper has a $\si$ and the helper's input and $\si$ is observed by the eavesdropper.
The rate-distortion-equivocation region is known for some special Markov chains among the $\si$ of nodes.
The authors have also studied the problem in which the helper does not receive any description but has access to a different $\si$, and the eavesdropper knows the output. 
If the output of the helper is available at the encoder, the region is known; otherwise, only for some special cases, it has been proven.
The problem of the triangular helper when the message to the helper and the receiver are the same has been studied in \cite{Bross16}.
In \cite{Benammar16ITW}, the source is made of two parts, and the channel between the encoder and the helper has infinite capacity but is observed by an eavesdropper.
There is no $\si$ at the receiver nor the eavesdropper, and the distortion is only dependent on the first source.

There are some approaches with an action-based encoder, such as \cite{Lu21}, in which $\si$s exist at the eavesdropper and the decoder. 
However, the $\si$s can be changed by an action taken by another decoder with some cost. 
In some cases, the whole region is found.
\fi
\makeatother

\makeatletter
\if@twocolumn
\else
\subsection{Contribution}
\fi
\makeatother
As the main contribution, we characterize the achievable rate-distortion-equivocation region of the setup in Fig. \ref{fig:system}, entirely.
We use two-level encoding similar to \cite{Villard13}.
The key is employed as a one-time pad, first protecting the second level of the encoded message, and then if key-rate is still available, the first part will be secured.
Interestingly, at the beginning of protecting the second part, the equivocation does not improve since the eavesdropper has $\si$, which is better for decoding that part.
Additionally, we show that from our framework, we can recover the lossless source coding, the setup with no shared-key, and the setup with no $\si$.
This setup has not been studied to the best of our knowledge, and the close studied setups are \cite{Villard13}, in which there is no shared-key between the encoder and the decoder; 
\cite{Chia13}, by assuming the same $\si$ between the sender and the receiver is the shared-key, so, there is no $\si$ at the decoder; 
\cite{Schieler14}, in which there is no $\si$ at the decoder and the notion of leakage is different; 
\cite{Merhav08}, by not assuming that the $\si$ at the eavesdropper is a degraded version of the $\si$ at the decoder, although the joint source-channel coding setup in \cite{Merhav08} is more general in other aspects;
and \cite{Kaspi15} in which the causal source coding and leakage have been considered, which results in a different perspective.

\makeatletter
\if@twocolumn
\else
\subsection{Organization}
\fi
\makeatother
This article is structured as follows.
We explain the notations used in this paper in the remainder of this section.
In Section \ref{sec:problem statement}, we define the problem formally.
In Section \ref{sec:main results}, we state our main result.
We identify our results for some special cases, such as lossless reconstruction, and compare them to the known results in Section \ref{sec:SpCases}.
Finally, we draw conclusions in Section \ref{sec:conclusions}.
\makeatletter
\if@twocolumn
The complete derivation of the proofs and more discussion can be found in the extended version of this paper in \cite{HPTM_ITW2021_long}.
\else
The proofs are in the appendix.
\fi
\makeatother

\makeatletter
\if@twocolumn
\emph{Notation}:
Sets, random variables ($\rv$s) and their realizations are denoted by calligraphic, capital and lower case letters, respectively.
The set of integer and real numbers are denoted by $\mathbb{N}$ and $\mathbb{R}$, respectively.
The set $\{1,\ldots,k\}$ for some $k\in\mathbb{N}$ is denoted by $[k]$.
The probability mass function ($\pmf$) of an $\rv$ $X$ with realizations $X=x$ defined on some alphabet $\mathcal{X}$ of finite cardinality $\lvert\mathcal{X}\rvert$ is denoted by $p_X(x)$ or just $p(x)$.
Similarly, for two r$\rv$s $X$ and $Y$, the conditional $\pmf$ of $Y$ given $X=x$ is denoted by $p_{Y|X}(y|x)$ or just $p(y|x)$.
The notations $\mathbb{E}[X]$, and $X \indep Y$ mean the expected value of $\rv$ $X$, and $X$ is independent of $Y$, respectively.
The sequence $(x_{m}, x_{m+1}, \ldots, x_n)$, for $m,n \in\mathbb{N}$, is denoted by $x_{m}^{n}$.
If $m=1$, we may use the notation $x^n$ instead of $x_1^n$.
All logarithms are in base $2$ unless otherwise stated.
\else
\subsection{Notation}
\begin{itemize}
	\item[-] Sets, random variables ($\rv$s) and their realizations are denoted by calligraphic, capital and lower case letters, respectively.
	The set of integer, rational and real numbers are denoted by $\mathbb{N}$, $\mathbb{Q}$ and $\mathbb{R}$, respectively.
	The set $\{1,\ldots,k\}$ for some $k\in\mathbb{N}$ is denoted by $[k]$.

	\item[-] The probability mass function ($\pmf$) of an $\rv$ $X$ with realizations $X=x$ defined on some alphabet $\mathcal{X}$ of finite cardinality $\lvert\mathcal{X}\rvert$ is denoted by $p_X(x)$ or just $p(x)$.
	Similarly, for two $\rv$s $X$ and $Y$, the conditional $\pmf$ of $Y$ given $X=x$ is denoted by $p_{Y|X}(y|x)$ or just $p(y|x)$.
	The notation $\mathbb{E}[X]$ means the expected value of $\rv$ $X$.
	The notation $X \indep Y$ means $X$ is independent of $Y$.
	
	\item[-]
	Random $\pmf$s are denoted by capital letter $P$.
	For instance, a random $\pmf$ for a $\rv$ $X$ with realization $x$ is denoted by $P_X(x)$, or just $P(x)$.
	Formally, it is equal to $p(x | B)$, where $B$ is a random object.
	The random object $B$ can be understood from the contex.
	In our framework, $B$ is always all the random binning functions in the system (note that it is the binning function, not the bin index).

	\item[-] The sequence $(x_{m}, x_{m+1}, \ldots, x_n)$, for $m,n \in\mathbb{N}$, is denoted by $x_{m}^{n}$.
	If $m=1$, we may use the notation $x^n$ instead of $x_1^n$.
	Also, $x^0$ means $\emptyset$.

	\item[-] $\mathbb{I}\{\mathcal{E}\}$ is a function of an event $\mathcal{E}$.
	It is equal to $1$ if the event $\mathcal{E}$ occures, and $0$ otherwise.

	\item[-] The norm $1$ distance between two $\pmf$s $p(x)$ and $q(x)$ is denoted by $\Vert p(x) - q(x)\Vert_1$, or just $\Vert p - q \Vert_1$ which is equal to $\sum_{x\in\mathcal{X}} |p(x) - q(x)|$.

	\item[-] All logarithms are in base $2$ unless otherwise stated.
	The term ``w.r.t.'' stands for ``with respect to''.
\end{itemize}
\fi
\makeatother

\section{Problem Statement} \label{sec:problem statement}
\begin{figure}
	\centering
	\makeatletter
	\if@twocolumn
		\includegraphics[trim = 0 40 0 50, clip, width=\columnwidth]{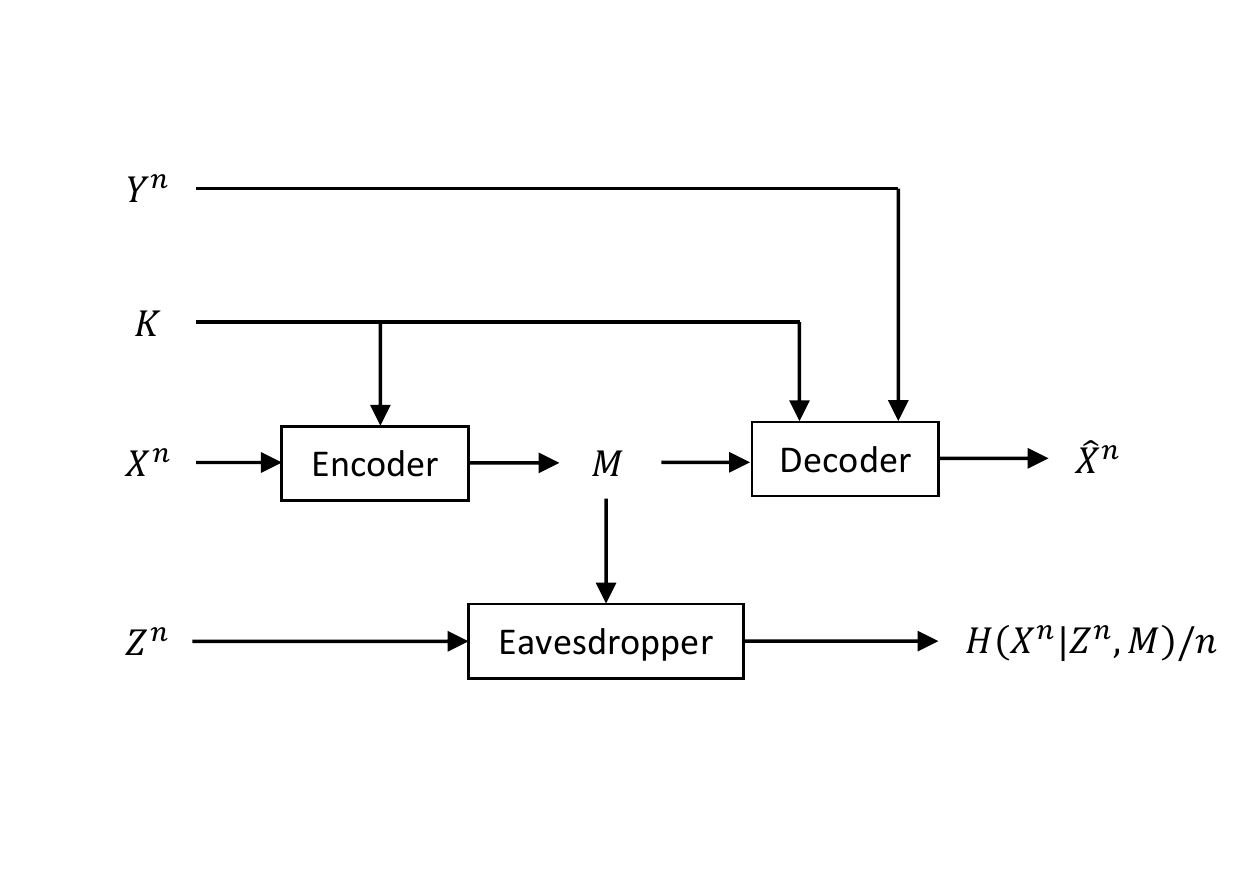}
	\else
		\includegraphics[trim = 0 40 0 50, clip, width=0.8\columnwidth]{problem}
	\fi
	\makeatother
	\caption{The sender encodes the source $X^n$ using the shared key $K$, and the receiver decodes the index $M$ utilizing the $\si$ $Y^n$ and $K$. 
	The eavesdropper has access to the index $M$ and $\si$ $Z^n$ and the equivocation is measured by $H(X^n \mid Z^n, M) / n$.}
	\label{fig:system}
\end{figure}

\begin{figure}
	\centering
	\makeatletter
	\if@twocolumn
		\includegraphics[trim = 0 40 0 50, clip, width=\columnwidth]{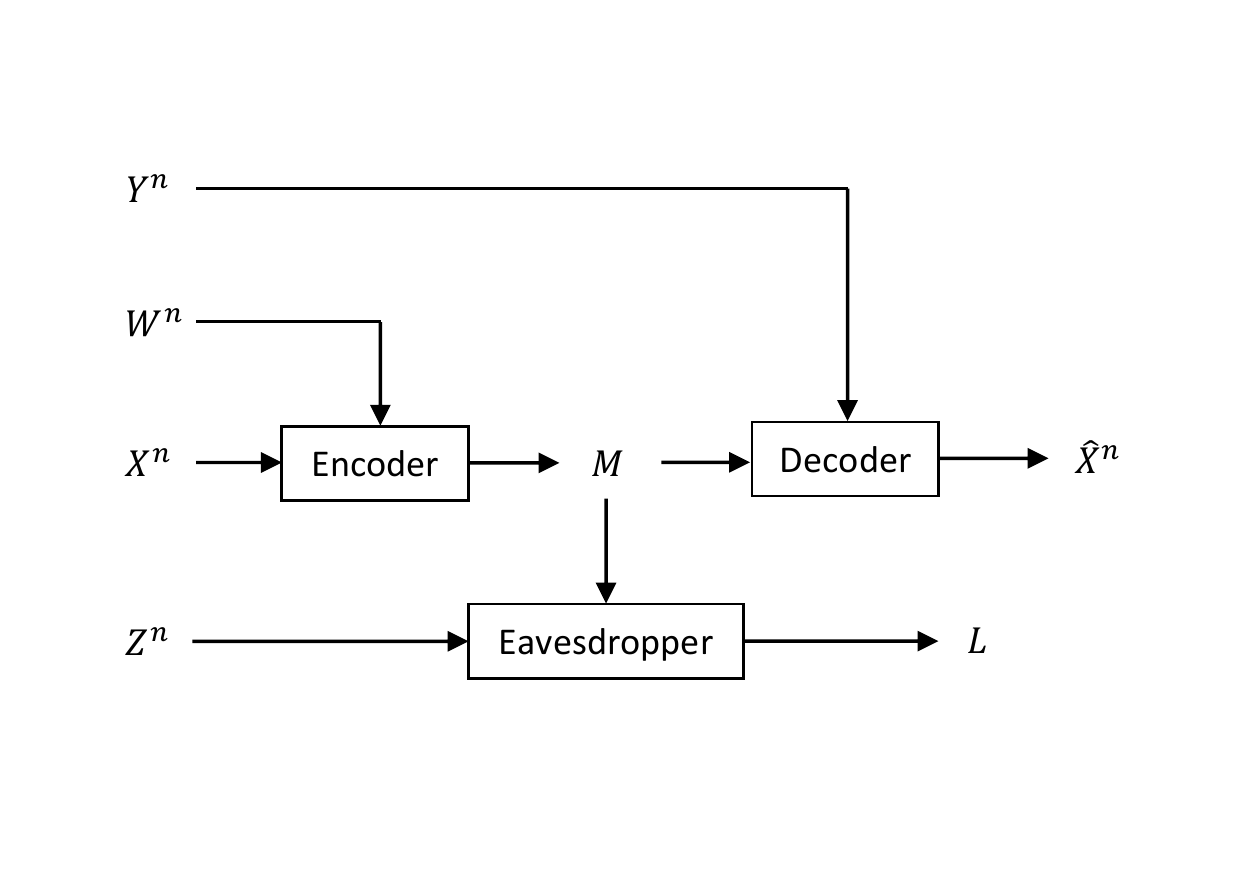}
	\else
		\includegraphics[trim = 0 40 0 50, clip, width=0.8\columnwidth]{genproblem}
	\fi
	\makeatother
	\caption{A general framework whose special case (Fig. \ref{fig:system}) is considered in this work.
	The variable $L$ is $H(X^n | Z^n,M )/n$ unless otherwise stated, and characterizes the normalized equivocation at the eavesdropper.}
	\label{fig:gensystem}
\end{figure}

Our setup is illustrated in Fig. \ref{fig:system}. 
In that setup, an independent key $K$ with rate $R_0$ is shared between the encoder and the decoder, whereas the decoder has additional access to some $\si$ $Y^n$. 
The eavesdropper has access to the $\si$ $Z^n$. 
The goal is to find the rate-distortion-leakage region of the source $X^n$.
Next, we formally define the problem.

\begin{definition}[Source Coding with Rate $R$ and Shared-key Rate $R_0$] \label{def:code}
	Assume $\left\{(X_i, Y_i, Z_i)\right\}_{i=1}^n$ is a block of a tuple of $\rv$s each one defined on the domain $\mathcal{X}\times\mathcal{Y}\times\mathcal{Z}$.
	Further, let $K$ be a shared-key with rate $R_0$, i.e., a $\rv$, independent of the sequence $\left\{(X_i, Y_i, Z_i)\right\}_{i=1}^n$, distributed uniformly on $\left\{1,\ldots,2^{\floor{n R_0}}\right\}$.
	A source code with rate $R$ and shared-key with rate $R_0$ consists of
	\begin{itemize}
		\item a (stochastic) encoder $\varphi$ that assigns a (possibly random) index with rate $R$ to the source and shared key, i.e.,
		\begin{equation*}
			\varphi \colon \mathcal{X}^{n}\times\left\{1,\ldots,2^{\floor{n R_0}}\right\} \to\mathcal{M},
			\quad (x^n, k) \mapsto m, 
		\end{equation*}
		where $\mathcal{M} := \left\{ 1, \ldots, 2^{\floor{n R}} \right\}$,
		\item a (stochastic) decoder, $\psi$, that (possibly randomly) reconstructs $\hat{X}^n$ based on the output of the encoder, the key, and the $\si$ at the decoder, i.e.,
		\makeatletter
		\if@twocolumn
			\begin{IEEEeqnarray*}{l}
				\psi\colon \mathcal{M}\times\left\{1,\ldots,2^{\floor{n R_0}}\right\}\times\mathcal{Y}^n \to \hat{\mathcal{X}}^n, \\
				(m, k, y^n) \mapsto \hat{x}^n,
			\end{IEEEeqnarray*}
		\else
			\begin{equation*}
				\psi\colon\mathcal{M}\times\left\{1,\ldots,2^{\floor{n R_0}}\right\}\times\mathcal{Y}^n \to \hat{\mathcal{X}}^n,
				\qquad (m, k, y^n) \mapsto\hat{x}^n,
			\end{equation*}
		\fi
		\makeatother
		where $\hat{\mathcal{X}}$ denotes the reconstruction domain.
	\end{itemize}
\end{definition}

Next, we define the achievability region. 
\begin{definition}[Achievable $(R, R_0, D, \Delta)$] \label{def:ach(R,D,delta)}
	Assume that $\left\{(X_i, Y_i, Z_i)\right\}_{i=1}^\infty$ is a sequence of $\rv$s, each with support $\mathcal{X}\times\mathcal{Y}\times\mathcal{Z}$.
	A tuple $(R, R_0, D, \Delta)$ is achievable with average normalized distortion level less than $D$ and equivocation greater than $\Delta$, if there exists a sequence of source codes with rate $R$, shared-key rate $R_0$, and output support $\hat{\mathcal{X}}$, such that
	\begin{IEEEeqnarray}{l}
		\limsup_{n\to\infty}{\E{d(X^n,\hat{X}^n)}}
		\leq D, \label{eqn:limsupEd<D}\\
		\liminf_{n\to\infty}\frac{1}{n}\Hb{X^n \mid M, Z^n} \geq \Delta,
		\quad \forall i\in\{1, 2, \ldots, k\}, \IEEEeqnarraynumspace\label{eqn:equiv>delta}
	\end{IEEEeqnarray}
	where for the given distortion function $d \colon \mathcal{X}\times\hat{\mathcal{X}} \to [0,\infty)$, we have
	\begin{equation*}
		d(X^n,\hat{X}^n)
		:= \frac{1}{n} \sum_{i=1}^{n}{d(X_i,\hat{X}_i)}.
	\end{equation*}
\end{definition}

\section{Main Results} \label{sec:main results}
The following theorem identifies the whole achievable region of $(R, R_0, D, \Delta)$.
\begin{theorem} \label{thm:R(D)}
	Consider the setup of Fig. \ref{fig:system}.
	Assume $p(x,y,z)$ is the $\pmf$ of the i.i.d. sequence with finite support $\mathcal{X}\times\mathcal{Y}\times\mathcal{Z}$.
	Then, $(R, R_0, D, \Delta)$ is achievable in the sense of Definition \ref{def:ach(R,D,delta)}, if and only if
	\makeatletter
	\if@twocolumn
		\begin{IEEEeqnarray}{rCl}
			R &\geq& \I{X}{V \mid Y}, \label{eqn:R(SI+SR)} \\
			\Delta &\leq& \min\{\I{Y}{V \mid U} - \I{Z}{V \mid U} + \H{X \mid Z, V} + R_0, \nonumber\\
			&&\quad\H{X \mid Z, U}\}, \label{eqn:Delta(SI+SR)} \\
			D &\geq& \E{d(X, \hat{X})}, \label{eqn:D(SI+SR)}
		\end{IEEEeqnarray}
	\else
		\begin{IEEEeqnarray}{rCl}
			R &\geq& \I{X}{V \mid Y}, \label{eqn:R(SI+SR)} \\
			\Delta &\leq& \min\{\I{Y}{V \mid U} - \I{Z}{V \mid U} + \H{X \mid Z, V} + R_0,
			\H{X \mid Z, U}\}, \label{eqn:Delta(SI+SR)} \\
			D &\geq& \E{d(X, \hat{X})}, \label{eqn:D(SI+SR)}
		\end{IEEEeqnarray}
	\fi
	\makeatother
	for some auxiliary $\rv$s $(U,V)$ with conditional $\pmf$ $p(v | x) \: p(u | v)$ over $\mathcal{U}\times\mathcal{V}$ and function $\hat{x}(y, v)$.
	Further, it is sufficient to have $|\mathcal{U}| \leq |\mathcal{X}|+4$ and $|\mathcal{V}|\leq (|\mathcal{X}|+3) (|\mathcal{X}|+4)$.
\end{theorem}

\begin{IEEEproof}
\makeatletter
\if@twocolumn
Due to space limitations, we only sketch the proof. 
The complete proof can be found in \cite[Theorem 1]{HPTM_ITW2021_long}.

\emph{Sketch of the proof of the achievability}:
Assume a tuple $(R,R_0,D,\Delta)$ satisfies \eqref{eqn:R(SI+SR)}, \eqref{eqn:Delta(SI+SR)}, and \eqref{eqn:D(SI+SR)}.
The basic idea of coding is similar to the idea of \cite{Villard13}, which we explain first.
Then we explain how to incorporate a shared key.

Due to allowed distortion, the encoder does not need to compress $X^n$ perfectly and compresses $V^n$ instead of $X^n$. 
As a result, $\I{X}{V}$ bits are needed. 
In addition, having $Y^n$ as $\si$ at the decoder makes it possible to further reduce by $\I{Y}{V}$ bits (Wyner-Ziv coding \cite[Theorem 11.3]{ElGamal11}).
This results in $\I{X}{V} - \I{X}{Y} = \I{X}{V \mid Y}$ bits for encoding.

Regarding security, using this technique, the eavesdropper would have a rate of $\I{Y}{V} - \I{Z}{V}$ bits uncertainty about $V^n$, and a rate of $\I{Y}{V} - \I{Z}{V} + \H{X \mid Z, V}$ bits about $X^n$. 
It has been shown in \cite{Villard13} that we can increase the uncertainty using another $\rv$ $U^n$ which has some information about $V^n$.

Intuitively, we want $U^n$ only to be dependent on the part of $V^n$ whose information can be obtained, by the eavesdropper, with fewer bits of description in comparison to the receiver, i.e., the mutual information between $U^n$ and $Z^n$ is greater than the one between $U^n$ and $Y^n$.
Hence, the sender, in the first step, encodes $U^n$; so the receiver will also have $U^n$, perfectly. 
Then, in the second step, the sender encodes $V^n$ assuming $U^n$ as a new $\si$ at the decoder.
In this case, the equivocation would become $\I{Y}{V \mid U} - \I{Z}{V \mid U} + \H{X\mid Z,V}$ using the fact that $\H{X\mid Z,U,V} = \H{X\mid Z,V}$ due to the Markov chain $U\to V\to (X,Y,Z)$.

Note that this does not change the necessary rate of encoding because, similar to the previous explanation, for the first step, $\I{X}{U}-\I{Y}{U}=\I{X}{U \mid Y}$ is sufficient, and for the second step, $\I{X}{V \mid U} - \I{Y}{V \mid U} = \I{X}{V \mid Y, U}$ is enough since $U^n$ is available at both the sender and the receiver.
So, the total rate is $\I{X}{U \mid Y} + \I{X}{V \mid Y, U} = \I{X}{U, V \mid Y} = \I{X}{V \mid Y}$ due to the Markov chain $U\to V\to (X,Y,Z)$.
However, this approach helps increase the equivocation since the eavesdropper may have a $\si$  which is better for decoding $U^n$.

In more detail, assume that $U^n$ can be decoded by the eavesdropper easier than the receiver due to the different $\si$.
Because the total amount of the necessary rate is fixed whether the sender encodes $V^n$ or $(U^n, V^n)$, the sender spends more rate to encode $U^n$, and encodes the remaining uncertainty of $V^n$, in the second step.
Hence, it causes the eavesdropper to have less information about the remaining uncertainty of $V^n$, other than $U^n$, which is known by the eavesdropper, perfectly.
Therefore, $U^n$ adds additional degrees of freedom to the problem, which makes the region larger.

\begin{figure}
	\centering
	\makeatletter
	\if@twocolumn
		\includegraphics[trim = 0 0 0 0, clip, width=\columnwidth]{intuition}
	\else
		\includegraphics[trim = 0 0 0 0, clip, width=0.7\columnwidth]{intuition}
	\fi
	\makeatother
	\caption{The equivocation with respect to key-rate diagram.
	It shows that when the key-rate increases, at some point increasing the rate does not increase the equivocation because of the strong correlation between the $\si$ at the eavesdropper $Z^n$ and $U^n$.
	Hence, $[\I{Z}{U} - \I{Y}{U}]^+ := \max\{0, \I{Z}{U} - \I{Y}{U}\}$ bits must be spent to compensate it.}
	\label{fig:insight}
\end{figure}

Now, we explain the case with the shared key.
For a better understanding, an illustration is shown in Fig. \ref{fig:insight}.
By having a shared key, first, the sender and decoder make $V^n$ secure, using the one-time pad technique \cite[Theorem 22.3]{ElGamal11}.
Hence, the equivocation increases with slope $1$ until all the information in the second part is perfectly secured (Line segment $1$ in Fig. \ref{fig:insight}).
This requires $R_0=\I{X}{V \mid Y,U}$ bits to secure all the bits used in the second step of encoding.

Then, if key-rate is still available, with the remaining bits of the shared key, they secure $U^n$.
However, the equivocation does not improve when the key-rate is not high enough.
In fact, to secure $U^n$, if the bits are fewer than $\I{Z}{U} - \I{Y}{U}$, the eavesdropper can still decode $U^n$, perfectly. 
Therefore, for the first $\max\{0,\I{Z}{U} - \I{Y}{U}\}$ bits, the equivocation will not be changed (Line segment $2$ in Fig. \ref{fig:insight}).
Afterwards, as the usage of bits increases, the equivocation also increases with solpe one (using one-time pad) until $U^n$ is secured (Line segment $3$ in Fig. \ref{fig:insight}) and the equivocation will remain $\H{X \mid Z}$ (Line segment $4$ in Fig. \ref{fig:insight}).

Nevertheless, this may look different compared with \eqref{eqn:Delta(SI+SR)}.
Note that the diagram coincides with the case when there is no $U$ in the encoding (see Fig. \ref{fig:insight}) in the last two parts.
Hence, in these cases, we can use the scheme with $U=\emptyset$.
Thus, it is sufficient only to consider the first two parts of the diagram (corresponding to the two constraints in \eqref{eqn:Delta(SI+SR)}).

\emph{Sketch of the proof of the converse}:
Assume that a tuple $(R,R_0,D,\Delta)$ is achievable in the sense of Definition \ref{def:ach(R,D,delta)}. 
The most important step is the following identification of the auxillary $\rv$s assignment, with conditional distribution $p(v | x) \: p(u | v) \: p(\hat{x} | v, y)$ such that they satisfy \eqref{eqn:R(SI+SR)}, \eqref{eqn:Delta(SI+SR)}, and \eqref{eqn:D(SI+SR)}.

Here, we just express the auxilary $\rv$s we used in the derivation.
For $i=1,\ldots,n$, we assign 
\begin{equation*}
	\begin{cases}
	U_i := \left(Y_{i+1}^n, Z^{i-1}, M\right), \\
	V_i := \left(X^{i-1}, Y^{i-1}, Y_{i+1}^n, Z^{i-1}, M, K \right), \\
	Q\sim\mathrm{Unif}\{1,\ldots,n\}, \\
	V := (V_Q, Q), \\
	Q := (U_Q, Q),
	\end{cases}
\end{equation*}
where $Q$ is independent of $(M, K, X^n, Y^n, Z^n, \hat{X}^n)$.
\begin{figure}
	\centering
	\makeatletter
	\if@twocolumn
		\includegraphics[trim = 0 20 0 30, clip, width=\columnwidth]{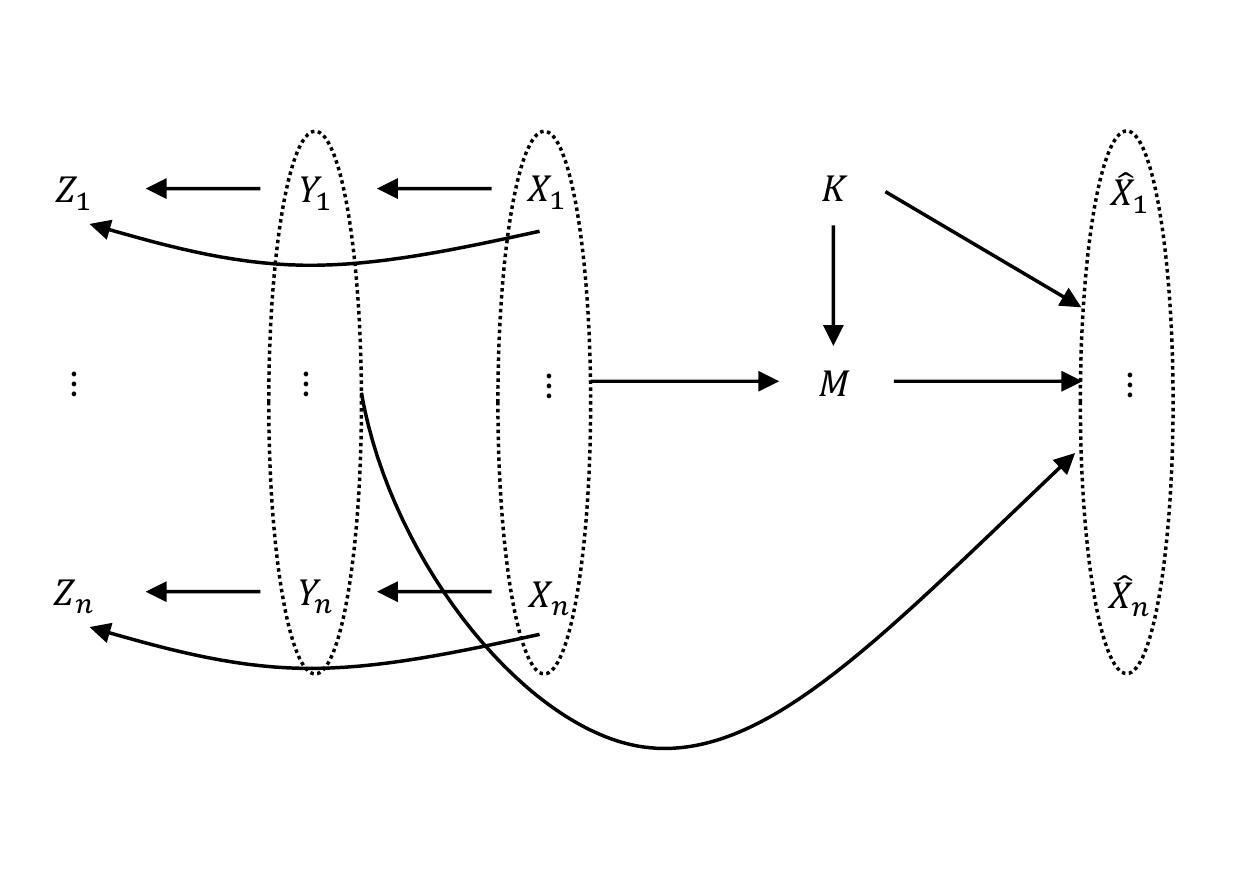}
	\else
		\includegraphics[trim = 0 20 0 30, clip, width=0.7\columnwidth]{converse1}
	\fi
	\makeatother
	\caption{Graphical representation of the main $\rv$s defined in the problem.}
	\label{fig:converse_sketch}
\end{figure}
From Fig. \ref{fig:converse_sketch}, by marginalizing over $\hat{X}^n$, it can be seen that we have the Makov chain $U_i \rightarrow V_i \rightarrow X_i \rightarrow (Y_i, Z_i)$, for $i=1,\ldots,n$.
For the remaining, we use the same conventional methods, and inequalities such as Csisz\'ar sum identity \cite[p. 25]{ElGamal11}, similar to \cite{Villard13}.
\else
	See the Appendix \ref{sec:Prf:Thm:RDE}.
\fi
\makeatother
\end{IEEEproof}

\section{Special Cases} \label{sec:SpCases}
This section finds the achievable region for some special cases and recovers known results in the literature.

\begin{corollary}[Lossless reconstruction] \label{cor:lossless}
	Assume that the distortion function is 
	\begin{equation*}
		d(x,\hat{x})=
		\begin{cases}
			0 & x=\hat{x},\\
			1 & x\neq\hat{x}.
		\end{cases}
	\end{equation*}
	The achievable tuples $(R, R_0, \Delta, 0)$ is
	\begin{IEEEeqnarray}{rCl}
		R &\geq& \H{X \mid Y}, \label{eqn:LL:R(SI+SR)} \\
		\Delta &\leq& \min\{\I{X}{Y \mid U} - \I{X}{Z \mid U} + R_0, \nonumber\\
		&&\qquad\H{X \mid Z, U}\}, \label{eqn:LL:Delta(SI+SR)}
	\end{IEEEeqnarray}
	for some conditional $\pmf$ $p(u | x)$.
\end{corollary}

\begin{IEEEproof}
	The proof of the achievability follows from Theorem \ref{thm:R(D)} by selecting $V=X$ and $\hat{x}(y,v) = v = x$; as a result, $D := \E{d(X,\hat{X})} = \Pr\{X\neq \hat{X}\} = 0$. \\
	For the proof of the converse, from $D=0$, we obtain $X=\hat{X}$; as a result, $\H{X \mid Y, V} = \H{\hat{X} \mid Y, V} = 0$, where it follows from the fact that $\hat{X}$ is a function of $(Y,V)$.
	Hence, \eqref{eqn:LL:R(SI+SR)} follows from \eqref{eqn:R(SI+SR)} and \eqref{eqn:LL:Delta(SI+SR)} follows from \eqref{eqn:Delta(SI+SR)} utilizing \eqref{eqn:VPForm} in Lemma \ref{lmm:=}.
\end{IEEEproof}

The following corollary shows the same result as \cite[Theorem 3]{Villard13}.
\begin{corollary}[$R_0 = 0$] \label{cor:K=0}
	If there is no shared-key ($R_0=0$), the achievable region is the same as Theorem \ref{thm:R(D)}, with substitution of \eqref{eqn:Delta(SI+SR)} with
	\begin{IEEEeqnarray*}{rCl}
		\Delta &\leq& \I{Y}{V \mid U} - \I{Z}{V \mid U} + \H{X \mid Z, V}.
	\end{IEEEeqnarray*}
\end{corollary}

\begin{IEEEproof}
From \eqref{eqn:SI+SR:Delta UP simple1} in Lemma \ref{lmm:=}, the right-hand side of \eqref{eqn:Delta(SI+SR)}, for $R_0 = 0$ is $\H{X \mid Z, U} - \I{X}{V \mid Y, U}$, which yields the result due to \eqref{eqn:VPForm} in Lemma \ref{lmm:=}.
\end{IEEEproof}

\begin{corollary}[$Y=\emptyset$] \label{cor:Y=0}
	If $Y=\emptyset$, the achievable region of tuple $(R,R_0,D,\Delta)$, $\mathcal{R}^*$, is
	\begin{IEEEeqnarray}{rCl}
		R &\geq& \I{X}{\hat{X}, U}, \label{eqn:Y=0:R(SI+SR)} \\
		\Delta &\leq& \min\{R_0 - \I{Z}{\hat{X} \mid U} + \H{X \mid Z, \hat{X}, U}, \nonumber\\
		&&\qquad\H{X \mid Z, U}\}, \label{eqn:Y=0:Delta(SI+SR)} \\
		D &\geq& \E{d(X, \hat{X})}. \label{eqn:Y=0:D(SI+SR)}
	\end{IEEEeqnarray}
	for some conditional $\pmf$ $p(\hat{x} , u | v)$ over $\hat{\mathcal{X}}\times\mathcal{U}$.
\end{corollary}

\begin{IEEEproof}
	The proof of the achievability follows from Theorem \ref{thm:R(D)} by selecting $V=(\hat{X}, U)$ and removing terms having $Y$. \\
	For the proof of the converse, from $Y=\emptyset$, we obtain $\hat{X}$ is a function of $V$; as a result, $\H{\hat{X} \mid V} = 0$.
	Hence, \eqref{eqn:LL:R(SI+SR)} follows from \eqref{eqn:R(SI+SR)} utilizing the fact that 
	\begin{equation*}
		\I{X}{V} \stackrel{(a)}{=} \I{X}{\hat{X}, V, U} \geq \I{X}{\hat{X},U},
	\end{equation*}
	where $(a)$ follows from the Markov chain $(\hat{X},U) \to V \to (X,Z)$.
	Similarly, \eqref{eqn:LL:Delta(SI+SR)} follows from \eqref{eqn:Delta(SI+SR)}, by selecting $Y=\emptyset$ and utilizing 
	 \begin{IEEEeqnarray*}{l}
	 	\I{Z}{V \mid U} \stackrel{(a)}{=} \I{Z}{V, \hat{X} \mid U} \geq \I{Z}{\hat{X} \mid U}, \\
	 	\H{X\mid Z, V} \stackrel{(b)}{=} \H{X \mid Z, V, \hat{X}, U} \leq \H{X \mid Z, \hat{X}, U},
	 \end{IEEEeqnarray*}
	 where $(a)$ and $(b)$ follow from the Markov chain $(\hat{X},U) \to V \to (X,Z)$.
\end{IEEEproof}

\begin{remark} \label{rmk:Ri<R<Ro}
	For $\mathcal{R}^*$, defined in Corollary \ref{cor:Y=0}, we have
	\begin{equation*}
		\mathcal{R}_\mathrm{in} \subseteq \mathcal{R}^* \subseteq \mathcal{R}_\mathrm{out},
	\end{equation*}
	where $\mathcal{R}_\mathrm{in}$ and $\mathcal{R}_\mathrm{out}$ are defined in \cite[Propositions 1 and 2]{Chia13} for $K'$ instead of $Y$ where $K' \indep (X,Z)$ and $\H{K'}=R_0$.
\end{remark}

\makeatletter
\if@twocolumn
\begin{IEEEproof}
\emph{Proof of $\mathcal{R}^* \subseteq \mathcal{R}_\mathrm{out}$}:
	From \cite[Propositions 1]{Chia13}, we obtain $\mathcal{R}_\mathrm{out}$ contains the tuples $(R,R_0,D,\Delta)$ such that
	\begin{IEEEeqnarray}{l}
		R \geq \I{X}{U',V' \mid K'}, \label{eqn:O:R(SI+SR)}\\
		\Delta \leq \min\{\H{X \mid Z}, 
		\H{X \mid Z, V', U'} + \I{K'}{V' \mid U'} \nonumber\\
		\qquad\qquad - \I{Z}{V' \mid U'} + \H{K' \mid U',V',X,Z} \}, \label{eqn:O:Delta(SI+SR)}\\
		D \geq \E{d(X, \hat{X}')} \label{eqn:O:D(SI+SR)},
	\end{IEEEeqnarray}
	for some conditional $\pmf$ $p(u',v' | x,k')$ and a function $\hat{x}'(u',v',k')$.
	Let $(U,\hat{X})$ satisfy conditions in Corollary \ref{cor:Y=0}.
	By selecting $U'=U$, $V'=(U,\hat{X})$, and $\hat{X}' = \hat{X}$, all of them independent of $K'$, \eqref{eqn:O:R(SI+SR)}, \eqref{eqn:O:Delta(SI+SR)}, and \eqref{eqn:O:D(SI+SR)} follow from \eqref{eqn:Y=0:R(SI+SR)}, \eqref{eqn:Y=0:Delta(SI+SR)}, and \eqref{eqn:Y=0:D(SI+SR)}, respectively, utilizing the facts that $\H{K'}=R_0$ and $\H{X\mid Z, U} \leq \H{X\mid Z}$.
	
\emph{Proof of $\mathcal{R}_\mathrm{in} \subseteq \mathcal{R}^*$}:
	From \cite[Propositions 2]{Chia13}, we obtain $\mathcal{R}_\mathrm{in}$ contains the tuples $(R,R_0,D,\Delta)$ such that
	\begin{IEEEeqnarray}{l}
		R > \I{X}{U',V' \mid K'}, \label{eqn:I:R(SI+SR)}\\
		\Delta < \min\{\H{X \mid Z, U'}, 
		\H{X \mid Z, U'} - \I{X}{V' \mid U', K'} \nonumber\\
		\qquad\qquad + \H{K' \mid U',V',X,Z} \}, \label{eqn:I:Delta(SI+SR)}\\
		D > \E{d(X, \hat{X}')}, \label{eqn:I:D(SI+SR)}
	\end{IEEEeqnarray}
	for some conditional $\pmf$ $p(u',v' | x,k')$ and a function $\hat{x}'(u',v',k')$.
	Let $(U',V',\hat{X}')$ satisfies the above inequalities of $\mathcal{R}_\mathrm{in}$.
	We select $U=U'$, $V=(U',V',K')$, and $\hat{x}(v) = \hat{x}'(u',v',k')$.
	Note that the Markov chain $U \to V \to X \to Z$ is satisfied due to $K\indep (X,Z)$.
	Hence, we obtain
	\begin{IEEEeqnarray*}{rCl}
		R &>& \I{X}{V \mid K'} \stackrel{(a)}{\geq} \I{X}{V} \\
		\Delta &<& \H{X \mid Z, U}, \\
		\Delta &<& \H{X \mid Z', U'} - \I{X}{V' \mid U', K'} \nonumber\\
		&& + \H{K' \mid U',V',X,Z} \\
		&=& \H{X \mid Z', U'} - \I{X}{V', K' \mid U'} \nonumber\\
		&& + \I{X}{K' \mid U'} + \H{K' \mid U',V',X,Z} \\
		&\leq& \H{X \mid Z', U'} - \I{X}{V', K' \mid U'} \nonumber\\
		&& + \I{X}{K' \mid U'} + \H{K' \mid U',X} \\
		&=& \H{X \mid Z', U'} - \I{X}{V', K' \mid U'}
		+ \H{K' \mid U'} \\
		&\stackrel{(b)}{\leq}& \H{X \mid Z', U'} - \I{X}{V', K' \mid U'} + R_0 \\
		&=& \H{X \mid Z, U} - \I{X}{V \mid U} + R_0 \\
		D &>& \E{d(X, \hat{X}')}
		= \E{d(X, \hat{X})}, \label{eqn:I:D(SI+SR)}
	\end{IEEEeqnarray*}
	where $(a)$ follows from $K' \indep X$;
	and $(b)$ follows from $\H{K'} = R_0$.
	Therefore, we obtain \eqref{eqn:R(SI+SR)}, \eqref{eqn:Delta(SI+SR)}, and \eqref{eqn:D(SI+SR)} for $Y=\emptyset$ utilizing \eqref{eqn:SI+SR:Delta UP simple1} in Lemma \ref{lmm:=}.
	It is the same as $\mathcal{R}^*$ as it is shown in Corollary \ref{cor:Y=0}.
\end{IEEEproof}
\else
\begin{IEEEproof}
	See Appendix \ref{sec:prf:rmk}.
\end{IEEEproof}
\fi
\makeatother

\begin{lemma} \label{lmm:=}
	\makeatletter
	\if@twocolumn
		\cite[Lemma 1]{HPTM_ITW2021_long}
	\fi
	\makeatother
	For $\rv$s $U, V, X, Y, Z$ with Markov chain $U\to V \to X \to (Y,Z)$, we have the following equalities:
	\makeatletter
	\if@twocolumn
		\begin{IEEEeqnarray}{rCl}
			\IEEEeqnarraymulticol{3}{l}{\I{Y}{V \mid U} - \I{Z}{V \mid U} + \H{X \mid Z, V}} \nonumber\\
			\quad &=& \H{X \mid Z} - \I{X}{V \mid Y}
			+ \I{Z}{U} - \I{Y}{U} \IEEEeqnarraynumspace \label{eqn:SI+SR:Delta UP simple2} \\
			&=& \H{X \mid Z, U} - \I{X}{V \mid Y, U} \label{eqn:SI+SR:Delta UP simple1} \\
			&=& \H{X \mid Y, V} + \I{X}{Y \mid U} - \I{X}{Z \mid U} \label{eqn:VPForm}.
		\end{IEEEeqnarray}
	\else
		\begin{IEEEeqnarray}{rCl}
			\IEEEeqnarraymulticol{3}{l}{\I{Y}{V \mid U} - \I{Z}{V \mid U} + \H{X \mid Z, V}} \nonumber\\
			\quad &=& \H{X \mid Y, V} + \I{X}{Y \mid U} - \I{X}{Z \mid U} \label{eqn:VPForm}\\
			&=& \H{X \mid Z, U} - \I{X}{V \mid Y, U} \label{eqn:SI+SR:Delta UP simple1}\\
			&=& \H{X \mid Z} - \I{X}{V \mid Y}
			+ \I{Z}{U} - \I{Y}{U}. \IEEEeqnarraynumspace \label{eqn:SI+SR:Delta UP simple2}
		\end{IEEEeqnarray}
	\fi
	\makeatother
\end{lemma}

\makeatletter
\if@twocolumn
\else
\begin{IEEEproof}
	See the Appendix \ref{prf:lmm:=}.
\end{IEEEproof}
\fi
\makeatother

\section{Conclusions} \label{sec:conclusions}
In this paper, we identified the achievable rate-distortion-equivocation region with side-information only available at the decoder, and a secure shared-key exists between the sender and receiver.
Further, the eavesdropper has access to the description of the source made by the encoder and $\si$ correlated to the source and $\si$ at the decoder.
The key is employed as a one-time pad, first protecting $V^n$ and then $U^n$ if key-rate is still available.
Interestingly, at the beginning of protecting $U^n$, the equivocation does not improve since the eavesdropper has $\si$ which is better for the decoding of $U^n$.
This problem serves as an intermediate step towards the more general problem when all parties in the system have different versions of $\si$, where we expect that a kind of key-extraction will be employed.

\bibliographystyle{IEEEtran}
\bibliography{myref}

\makeatletter
\if@twocolumn
\else
\appendices
\section{Proof of Theorem \ref{thm:R(D)}} \label{sec:Prf:Thm:RDE}
The derivation of the proof consists of two parts, the achievability and the converse.
In the achievability part, we show that if a tuple $(R_, R_0, D, \Delta)$ satisfies \eqref{eqn:R(SI+SR)}, \eqref{eqn:Delta(SI+SR)}, and \eqref{eqn:D(SI+SR)}, then it is achievable in the sense of Definition \ref{def:ach(R,D,delta)}.
In the converse part, we show that if a tuple $(R_, R_0, D, \Delta)$ is achievable in the sense of Definition \ref{def:ach(R,D,delta)}, then it satisfies \eqref{eqn:R(SI+SR)}, \eqref{eqn:Delta(SI+SR)}, and \eqref{eqn:D(SI+SR)}.

\subsection{Achievability}
	\sloppy Assume that the distribution $(X, Y, Z, \hat{X}, U, V) \sim p(x,y,z) \: p(\hat{x} | y,v) \: p(v | x) \: p(u | v)$ is given such that it satisfies \eqref{eqn:R(SI+SR)}, \eqref{eqn:Delta(SI+SR)}, and \eqref{eqn:D(SI+SR)}.
	To show that \eqref{eqn:limsupEd<D} and \eqref{eqn:equiv>delta} are correct, we use the method proposed in \cite{Yassaee14}.

\begin{figure}
	\begin{subfigure}{0.5 \textwidth}
		\centering
		\includegraphics[trim = 10 30 15 30, clip, width=\columnwidth]{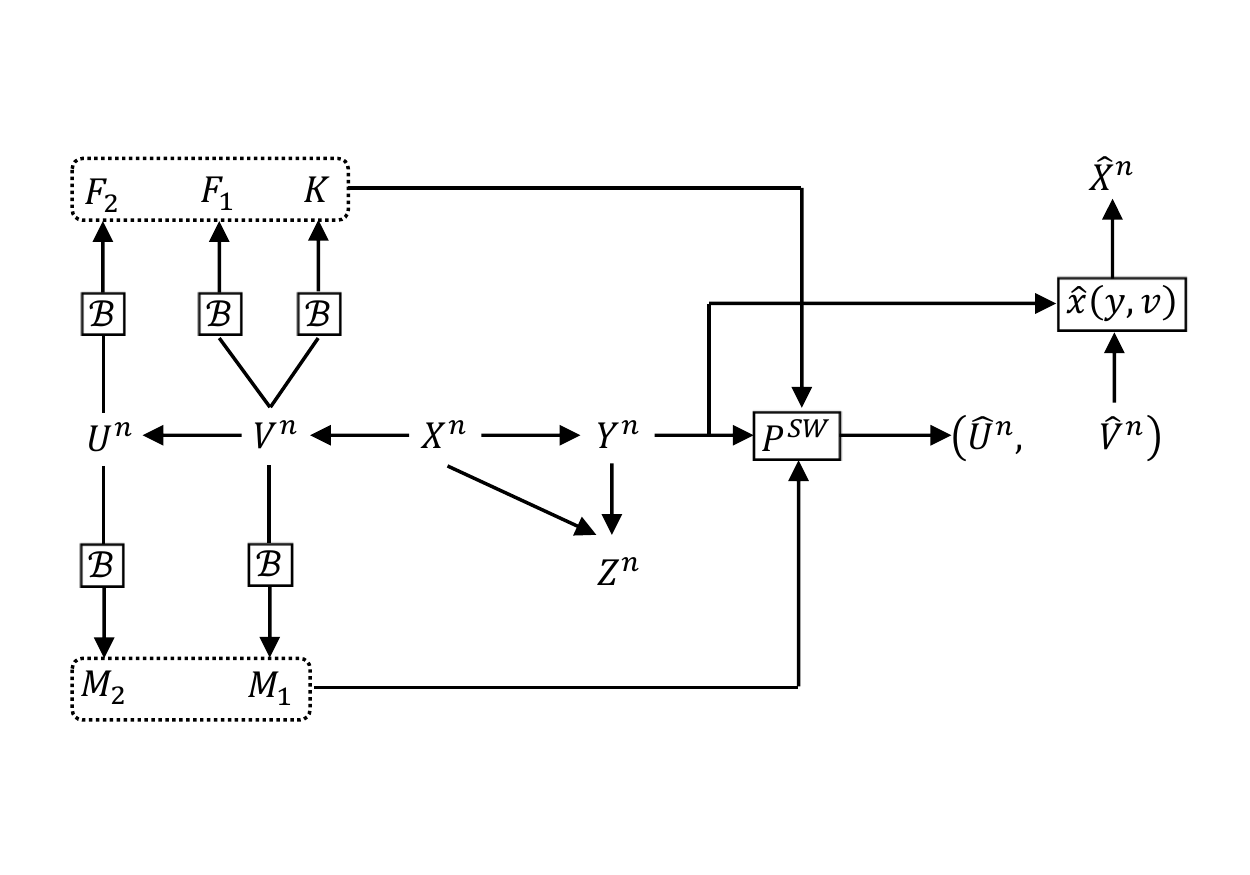}
		\caption{Protocol A}
		\label{fig:ProtocolA}
	\end{subfigure}
	\hfill
	\begin{subfigure}{0.5 \textwidth}
		\centering
		\includegraphics[trim = 0 30 0 30, clip, width=\columnwidth]{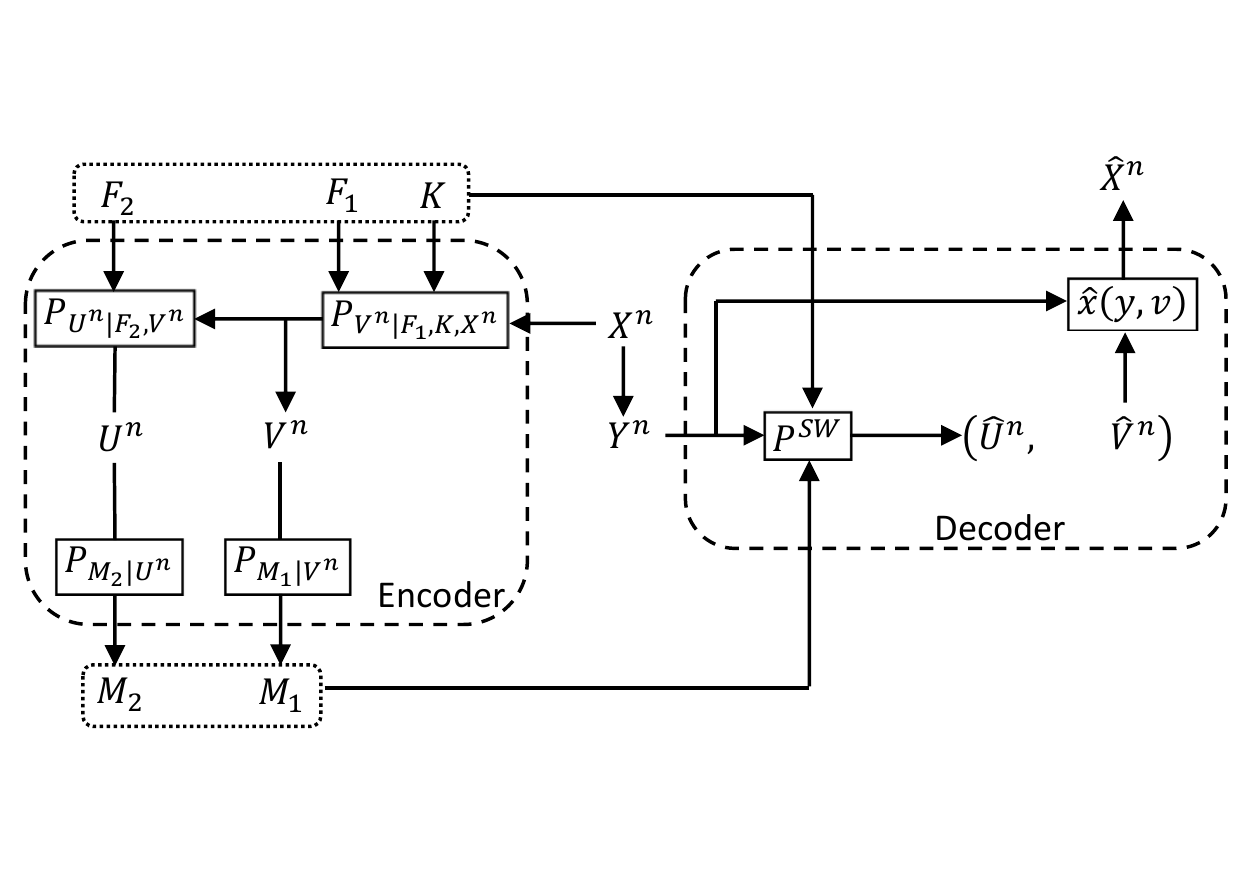}
		\caption{Protocol B}
		\label{fig:ProtocolB}
	\end{subfigure}
	\caption{Protocols A and B in the achivability proof of Theorem \ref{thm:R(D)}. 
	It is proved that if some constraints are satisfied, the joint probability of the $\rv$s in both protocols become equal with high probability. 
	Hence, the encoder and decoder can be defined from the joint probabilities of the first protocol.}
	\label{fig:achievability}
\end{figure}

	\noindent\textbf{Step 1}: Introducing Protocols A and B.\\
	We define two protocols each of which defines a distribution on the $\rv$s.
	
	\emph{Protocol A} (well-behaved distribution):
	Let $(X^n, Y^n, Z^n, U^n, V^n)$ be i.i.d. with $\pmf$ $p(x,y,z) \: p(v | x) \: p(u | v)$.
	As illustrated in Fig. \ref{fig:ProtocolA}, we do the following random binnings (all the random binnings are jointly independent and uniformly distributed over their domain):\\
	$M_1 \in [2^{n R_1}]$, $F_1 \in [2^{n \tilde{R}_1}]$, and $K \in [2^{n R_0}]$ are the binnings of  $V^n$; and
	$M_2 \in [2^{n R_2}]$ and $F_2 \in [2^{n \tilde{R}_2}]$ are the binnings of $U^n$.
	
	To obtain $(\hat{U}^n, \hat{V}^n)$, we utilize the Slepian-Wolf decoder \cite[Lemma 1]{Yassaee14} $(y^n, m_1, m_2, f_1, f_2, k) \mapsto (\hat{u}^n, \hat{v}^n)$ with induced distributions $P(\hat{u}^n, \hat{v}^n | y^n, m_1, m_2, f_1, f_2, k)$.
	Finally, to obtain $\hat{X}^n$, we have $\hat{X}_i=\hat{x}(Y_i, \hat{V}_i)$ for $i=1,\ldots,n$.
	Hence, the distribution is
	\begin{IEEEeqnarray}{rCl}
		\IEEEeqnarraymulticol{3}{l}{P_A(x^n, y^n, z^n, u^n, v^n, \hat{u}^n, \hat{v}^n, \hat{x}^n, m_1, m_2, f_1, f_2, k)} \nonumber\\
		\qquad &=& p(x^n, y^n, z^n) p(v^n | x^n) p(u^n | v^n) \nonumber\\
		&&\times P(f_1 | v^n) P(m_1 | v^n)  P(k | v^n) P(f_2 | u^n) P(m_2 | u^n) \nonumber\\
		&& \times P^{SW}(\hat{u}^n, \hat{v}^n | y^n, m_1, m_2, f_1, f_2, k) 
		p(\hat{x}^n| y^n, \hat{v}^n) \nonumber\\
		&=& p(x^n, y^n, z^n)
		P(f_1, k | x^n) P(v^n | x^n, f_1, k) P(m_1 | v^n) \nonumber\\
		&&\times P(f_2 | v^n) P(u^n | v^n, f_2) P(m_2 | u^n) \nonumber\\
		&& \times P^{SW}(\hat{u}^n, \hat{v}^n | y^n, m_1, m_2, f_1, f_2, k) 
		p(\hat{x}^n| y^n, \hat{v}^n), \IEEEeqnarraynumspace\label{eqn:PA}
	\end{IEEEeqnarray}
	where 
	\begin{equation*}
		p(\hat{x}^n|y^n, \hat{v}^n) = \prod_{i=1}^n\mathbb{I}\{x_i = \hat{x}(y_i, \hat{v}_i)\}.
	\end{equation*}
	
	\emph{Protocol B} (practical distribution):
	This protocol is similar to our problem statement. 
	We propose an encoder and a decoder for our problem, based on the induced distributions in Protocol A.
	$(X^n, Y^n, Z^n)$ are generated i.i.d. according to the distribution $p(x,y,z)$.
	We assume that $F_1 \sim\mathrm{Unif}[2^{n\tilde{R}_1}]$, $F_2 \sim\mathrm{Unif}[2^{n\tilde{R}_2}]$, $K \sim\mathrm{Unif}[2^{nR_0}]$, and $(F_1, F_2, K)$ are jointly independent.
	Further, the encoder, the decoder, and the eavesdropper have access to $F_1$ and $F_2$, while $K$ is only accessible by the encoder and the decoder.
	
	\textit{Encoder}: 
	The encoder, first, generates $(U^n, V^n)$ from $(X^n, F_1, F_2, K)$ via the distribution $P(v^n | x^n, f_1, k) P(u^n | v^n, f_2)$ introduced in Protocol A.
	Then, the encoder sends $(M_1, M_2)$ according to the distribution $P(m_1 | v^n) P(m_2 | u^n)$ introduced in Protocol A.
	
	\textit{Decoder}:
	The decoder finds $(\hat{U}^n, \hat{V}^n)$ from $(Y^n, M_1, M_2, F_1, F_2, K)$ using the Slepian-Wolf decoder $P^{SW}(\hat{u}^n, \hat{v}^n | y^n, m_1, m_2, f_1, f_2, k)$.
	Finally, $\hat{X}^n$ is generated similar to Protocol A, as the $\hat{X}_i=\hat{x}(Y_i, \hat{V}_i)$ for $i=1,\ldots,n$.
	
	Hence, the distribution is
	\begin{IEEEeqnarray}{rCl}
		\IEEEeqnarraymulticol{3}{l}{P_B(x^n, y^n, z^n, u^n, v^n, \hat{u}^n, \hat{v}^n, \hat{x}^n, m_1, m_2, f_1, f_2, k)} \nonumber\\
		\quad&=& p(x^n, y^n, z^n)
		2^{-n\tilde{R}_1} 2^{-nR_0} P(v^n | x^n, f_1, k) P(m_1 | v^n) \nonumber\\
		&&\times 2^{-n\tilde{R}_2} P(u^n | f_2, v^n) P(m_2 | u^n) \nonumber\\
		&& \times P^{SW}(\hat{u}^n, \hat{v}^n | y^n, m_1, m_2, f_1, f_2, k) 
		p(\hat{x}^n| y^n, \hat{v}^n). \label{eqn:PB} \IEEEeqnarraynumspace
	\end{IEEEeqnarray}
	
	\noindent\textbf{Step 2}: Making $P_A$ and $P_B$ identical, and the Slepian-Wolf decoder succeed,  asymptotically with high probability in norm $1$ distance.
	
	We can write the norm $1$ distance as following:
	\makeatletter
	\if@twocolumn
		\begin{IEEEeqnarray}{rCl}
			\IEEEeqnarraymulticol{3}{l}{\left\lVert P_A - P_B \right\rVert_1} \nonumber\\
			&=& \lVert \left[ P_A(x^n, y^n, z^n, f_1, f_2, k) - P_B(x^n, y^n, z^n, f_1, f_2, k) \right] \nonumber\\
			&& \quad \times P(u^n, v^n, m_1, m_2, \hat{u}^n, \hat{v}^n | x^n, y^n, z^n, f_1, f_2, k) \rVert_1 \nonumber\\
			&\stackrel{(a)}{=}& \lVert P_A(x^n, y^n, z^n, f_1, f_2, k) - P_B(x^n, y^n, z^n, f_1, f_2, k) \rVert_1, \label{eqn:|PA-PB|} \IEEEeqnarraynumspace
		\end{IEEEeqnarray}
	\else
		\begin{IEEEeqnarray}{rCl}
			\left\lVert P_A - P_B \right\rVert_1
			&=& \lVert \left[ P_A(x^n, y^n, z^n, f_1, f_2, k) - P_B(x^n, y^n, z^n, f_1, f_2, k) \right] \nonumber\\
			&& \quad \times P(u^n, v^n, m_1, m_2, \hat{u}^n, \hat{v}^n | x^n, y^n, z^n, f_1, f_2, k) \rVert_1 \nonumber\\
			&\stackrel{(a)}{=}& \lVert P_A(x^n, y^n, z^n, f_1, f_2, k) - P_B(x^n, y^n, z^n, f_1, f_2, k) \rVert_1, \label{eqn:|PA-PB|} \IEEEeqnarraynumspace
		\end{IEEEeqnarray}
	\fi
	\makeatother
	where $P(u^n, v^n, m_1, m_2, \hat{u}^n, \hat{v}^n | x^n, y^n, z^n, f_1, f_2, k)$ is the part which is equal in both $P_A$ and $P_B$;
	and $(a)$ follows from \cite[Lemma 3.1]{Yassaee14}.
	Hence, it is sufficient to have
	\makeatletter
	\if@twocolumn
		\begin{IEEEeqnarray}{l}
			\lim_{n\to\infty}\mathbb{E}\big[\Vert P_A(x^n, y^n, z^n, f_1, f_2, k) \nonumber\\
			\qquad - p(x^n, y^n, z^n) 2^{-n R_0} 2^{-n\tilde{R}_1} 2^{-n\tilde{R}_2} \Vert_1\big]
			= 0. \label{eqn:(Xn,F1)(Vn,F2) indep}
		\end{IEEEeqnarray}
	\else
		\begin{equation} \label{eqn:(Xn,F1)(Vn,F2) indep}
			\lim_{n\to\infty}\E{\Vert P_A(x^n, y^n, z^n, f_1, f_2, k)
			- p(x^n, y^n, z^n) 2^{-n R_0} 2^{-n\tilde{R}_1} 2^{-n\tilde{R}_2} \Vert_1}
			= 0.
		\end{equation}
	\fi
	\makeatother
	In order to obtain \eqref{eqn:(Xn,F1)(Vn,F2) indep}, from \cite[Theorem 1]{Yassaee14} (for $X_i \gets V$, for $i=1,2$, $X_i \gets U$, for $i=3$, and $Z\gets (X,Y,Z)$), it is sufficient to have
	\begin{IEEEeqnarray}{l}
		R_0 + \tilde{R}_1 < \H{V \mid X,Y,Z} = \H{V \mid X}, \label{eqn:Rf1,Rf2<1} \\
		\tilde{R}_2 < \H{U \mid X,Y,Z} = \H{U \mid X}, \label{eqn:Rf1,Rf2<2} \\
		R_0 + \tilde{R}_1 + \tilde{R}_2 < \H{U, V \mid X,Y,Z} = \H{U, V \mid X}, \label{eqn:Rf1,Rf2<3} \IEEEeqnarraynumspace
	\end{IEEEeqnarray}
	where the equalities follow from the Markov chain $(U,V) \to X \to (Y,Z)$ in Protocol A. \\
	Now, we need to find the conditions making the Slepian-Wolf decoder work properly, i.e.,
	\makeatletter
	\if@twocolumn
		\begin{IEEEeqnarray}{l}
			\lim_{n\to\infty}\mathbb{E}\big[\Vert P_A(y^n, u^n, v^n, \hat{u}^n, \hat{v}^n) \nonumber\\
			\qquad - \mathbb{I}\{u^n = \hat{u}^n, v^n = \hat{v}^n\} p(y^n, u^n, v^n) \Vert_1\big] 
			= 0. \label{eqn:SW works} \IEEEeqnarraynumspace
		\end{IEEEeqnarray}
	\else
		\begin{equation} \label{eqn:SW works}
			\lim_{n\to\infty}\E{\Vert P_A(y^n, u^n, v^n, \hat{u}^n, \hat{v}^n)
			- \mathbb{I}\{u^n = \hat{u}^n, v^n = \hat{v}^n\} p(y^n, u^n, v^n) \Vert_1} 
			= 0.
		\end{equation}
	\fi
	\makeatother
	From \cite[Lemma 1]{Yassaee14} (for $X_i \gets V$, for $i=1,2,3$, $X_i \gets U$, for $i=4,5$, and $Z\gets Y$), \eqref{eqn:SW works} follows if
	\begin{IEEEeqnarray}{l}
			R_0 + R_1 + \tilde{R}_1 > \H{V \mid Y, U}, \label{eqn:R+Rf >1} \\
			R_2 + \tilde{R}_2 > \H{U \mid Y, V} = \H{U \mid V}, \label{eqn:R+Rf >2} \\
			R_0 + R_1 + R_2 + \tilde{R}_1 + \tilde{R}_2 > \H{U, V \mid Y}, \label{eqn:R+Rf >3}
	\end{IEEEeqnarray}
	where the equality follows from the Markov chain $U \to V \to Y$ in Protocol A. \\
	Next, we try to make the coding secure.
	As we show later, it is sufficient to have
	\makeatletter
	\if@twocolumn
		\begin{IEEEeqnarray}{l}
			\lim_{n\to\infty}\mathbb{E}\big[\Vert P_A(x^n, z^n, u^n, m_1, f_1) \nonumber\\
			\qquad - p(x^n, z^n, u^n) 2^{-n R_1} 2^{-n\tilde{R}_1} \Vert_1 \big] 
			= 0. \label{eqn:Z indep}
		\end{IEEEeqnarray}
	\else
		\begin{equation} \label{eqn:Z indep}
			\lim_{n\to\infty}\E{\Vert P_A(x^n, z^n, u^n, m_1, f_1) 
			- p(x^n, z^n, u^n) 2^{-n R_1} 2^{-n\tilde{R}_1} \Vert_1} 
			= 0.
		\end{equation}
	\fi
	\makeatother
	From \cite[Theorem 1]{Yassaee14} (for $X_i \gets V$, for $i=1,2$, and $Z \gets (X,Z,U)$), \eqref{eqn:Z indep} it is concluded if 
	\begin{equation} \label{eqn:R+Rf < Z indep}
		R_1 + \tilde{R}_1 < \H{V \mid X, Z, U} = \H{V \mid X, U},
	\end{equation}
	where the equality follows from the Markov chain $(U,V)\to X \to Z$, in Protocol A.
	Later, utilizing Fourier-Motzkin elimination \cite[Appendix D]{ElGamal11}, we obtain that \eqref{eqn:Rf1,Rf2<1}, \eqref{eqn:Rf1,Rf2<2}, \eqref{eqn:Rf1,Rf2<3}, \eqref{eqn:R+Rf >1}, \eqref{eqn:R+Rf >2}, \eqref{eqn:R+Rf >3}, and \eqref{eqn:R+Rf < Z indep} are equivalent to
	\begin{equation} \label{eqn:Final ineq}
		\begin{cases}
			R_0 > \I{X}{V \mid Y, U}, \\
			R > \I{X}{V \mid Y}.
		\end{cases}
	\end{equation}

	Therefore, from \eqref{eqn:|PA-PB|}, \eqref{eqn:(Xn,F1)(Vn,F2) indep}, \eqref{eqn:SW works}, and \eqref{eqn:Z indep}, we obtain that, for all $\epsilon>0$, if \eqref{eqn:Final ineq} is satisfied, there exist binning functions, $b$, such that, for large enough $n$,
	\makeatletter
	\if@twocolumn
		\begin{IEEEeqnarray}{l}
			\lVert p_B(\cdot \mid b) - p_C(\cdot \mid b) \rVert_1
			\leq\epsilon, \label{eqn:PB-PC=0} \\
			\Vert p_B(x^n, z^n, u^n, m_1, f_1 \mid b)
			 - p(x^n, z^n, u^n) 2^{-n R_1} 2^{-n\tilde{R}_1} \Vert_1 \nonumber\\
			\qquad\leq\epsilon, \label{eqn:PB approx equivocation}
		\end{IEEEeqnarray}
	\else
		\begin{IEEEeqnarray}{l}
			\lVert p_B(\cdot \mid b) - p_C(\cdot \mid b) \rVert_1
			\leq\epsilon, \label{eqn:PB-PC=0} \\
			\Vert p_B(x^n, z^n, u^n, m_1, f_1 \mid b)
			- p(x^n, z^n, u^n) 2^{-n R_1} 2^{-n\tilde{R}_1} \Vert_1
			\leq\epsilon, \label{eqn:PB approx equivocation}
		\end{IEEEeqnarray}
	\fi
	\makeatother
	where
	\begin{IEEEeqnarray}{rCl}
		\IEEEeqnarraymulticol{3}{l}{p_C(x^n, y^n, z^n, u^n, v^n, \hat{u}^n, \hat{v}^n, \hat{x}^n, m_1, m_2, f_1, f_2, k \mid b)} \nonumber\\
		\quad &:=& \prod_{i=1}^n p(x_i, y_i, z_i) p(v_i | x_i) p(u_i | v_i) \nonumber\\
		&&\times p(f_1 | v^n, b) p(m_1 | v^n, b) p(k | v^n, b) \nonumber\\
		&&\times p(f_2 | u^n, b) p(m_2 | u^n, b) \nonumber\\
		&&\times\mathbb{I}\{u^n = \hat{u}^n, v^n = \hat{v}^n\}
		\prod_{i=1}^n p(\hat{x}_i|y_i, \hat{v}_i). \label{eqn:PC}
	\end{IEEEeqnarray}

	\noindent\textbf{Step 3}: Removing the common randomness $F_1, F_2$.
	
	To be able to provide feasible encoder and decoder, we cannot use distribution $P_B$ because they are a function of common randomness $(F_1, F_2)$ which does not exist in reality.
	Hence, it suffices to find an instance $(f_1, f_2)$ of $(F_1, F_2)$ such that the distribution $p(x^n, y^n, z^n)$ is not changed, and \eqref{eqn:limsupEd<D} and \eqref{eqn:equiv>delta} are satisfied.
	Then, $p(m_1,m_2|x^n,f_1,f_2,k,b)$ and $p^{SW}(\hat{x}^n|y^n, f_1,f_2,k,b)$ are the final encoder and decoder.
	
	We consider the following cases:
	\begin{itemize}
		\item $R_0 > \I{X}{V \mid Y, U}$:
		For the distortion, from \eqref{eqn:PC}, we have that for any small $\zeta>0$
		\begin{IEEEeqnarray*}{l}
			\mathbb{E}_{p_C}\big[\mathbb{E}_{p_C}\big[\mathbb{I}\{d(X^n, \hat{X}^n)>D+\zeta\} \mid F_1, F_2, b \big]\big] \\
			\qquad= \mathbb{E}_{p_C}\big[\mathbb{I}\{d(X^n, \hat{X}^n)>D+\zeta\} \mid b \big] \\
			\qquad= \mathrm{Pr}_{p_C}\{d(X^n, \hat{X}^n)>D+\zeta\} \\
			\qquad \stackrel{(a)}{\leq} \frac{d_\mathrm{max}^2}{n\left(D+\zeta-\E{d(X,\hat{X})}\right)^2},
		\end{IEEEeqnarray*}
		where $\mathbb{E}_{p_C}$ denotes the expectation with distribution $p_C$, 
		$d_\mathrm{max} := \max_{x,\hat{x}}{d(x,\hat{x})}$,
		and $(a)$ follows from \eqref{eqn:D(SI+SR)} and the Chebyshev's bound.
		Hence, from \cite[Lemma 5]{Yassaee14} and \eqref{eqn:PB-PC=0}, we have
		\makeatletter
		\if@twocolumn
			\begin{IEEEeqnarray}{rCl}
				\IEEEeqnarraymulticol{3}{l}{\mathbb{E}_{p_B}\big[\mathbb{E}_{p_B}\big[\mathbb{I}\{d(X^n, \hat{X}^n)>D+\zeta\} \mid F_1,F_2,b \big]\big]} \nonumber\\
				\quad&=& \mathrm{Pr}_{p_B}\{d(X^n, \hat{X}^n)>D+\zeta\} \nonumber\\
				&\leq& \epsilon d_\mathrm{max} + \frac{d_\mathrm{max}^2}{n\left(D+\zeta-\E{d(X,\hat{X})}\right)^2}. \label{eqn:d-d<e}
			\end{IEEEeqnarray}
		\else
			\begin{IEEEeqnarray}{rCl}
				\IEEEeqnarraymulticol{3}{l}{\mathbb{E}_{p_B}\big[\mathbb{E}_{p_B}\big[\mathbb{I}\{d(X^n, \hat{X}^n)>D+\zeta\} \mid F_1,F_2,b \big]\big]} \nonumber\\
				\qquad&=& \mathrm{Pr}_{p_B}\{d(X^n, \hat{X}^n)>D+\zeta\} \nonumber\\
				&\leq& \epsilon d_\mathrm{max} + \frac{d_\mathrm{max}^2}{n\left(D+\zeta-\E{d(X,\hat{X})}\right)^2}. \label{eqn:d-d<e}
			\end{IEEEeqnarray}
		\fi
		\makeatother
		For the equivocation, for the distribution $p_B$, we have
		\makeatletter
		\if@twocolumn
			\begin{IEEEeqnarray}{rCl}
				\IEEEeqnarraymulticol{3}{l}{\H{X^n \mid Z^n, M_1, M_2, F_1, F_2, b}} \nonumber\\
				\qquad&\stackrel{(a)}{\geq}& \H{X^n \mid Z^n, U^n, M_1, F_1, b} \nonumber\\
				&\stackrel{(b)}{=}& \H{X^n \mid Z^n, U^n, b} - n\delta \nonumber\\
				&\stackrel{(c)}{=}& \H{X^n \mid Z^n, U^n} - n\delta, \label{eqn:Delta case1}
			\end{IEEEeqnarray}
		\else
			\begin{IEEEeqnarray}{rCl}
				\H{X^n \mid Z^n, M_1, M_2, F_1, F_2, b}
				&\stackrel{(a)}{\geq}& \H{X^n \mid Z^n, U^n, M_1, F_1, b} \nonumber\\
				&\stackrel{(b)}{=}& \H{X^n \mid Z^n, U^n, b} - n\delta \nonumber\\
				&\stackrel{(c)}{=}& \H{X^n \mid Z^n, U^n} - n\delta, \label{eqn:Delta case1}
			\end{IEEEeqnarray}
		\fi
		\makeatother
		where $(a)$ follows due to the fact that, based on \eqref{eqn:PB}, $(M_2, F_2)$ is a function of $(U^n, b)$;
		$(b)$ follows from \eqref{eqn:PB approx equivocation} and \cite[Theorem 17.3.3]{Cover06} for  some $\delta \leq 2c\epsilon-2\epsilon\log\epsilon$ for $c=\log(|\mathcal{X}| |\mathcal{Z}| |\mathcal{U}|) + R$, which $\delta\to 0$ as $\epsilon\to 0$;
		and $(c)$ follows from \eqref{eqn:PB approx equivocation}.
		Thus, there exist $f_1$ and $f_2$ such that \eqref{eqn:limsupEd<D} and \eqref{eqn:equiv>delta} are satisfied.

		\item $R_0 \leq \I{X}{V \mid Y, U}$:
		We assume that there is a shared-key $\bar{K}$ with rate $\bar{R}_0 = \I{X}{V \mid Y, U} + \zeta$, for some small $\zeta>0$.
		Then, $\bar{K}$ is split into two independent parts $K$ and $K'$ with rates $R_0$ and $\bar{R}_0 - R_0$, respectively.
		We claim that the problem is equivalent to the case when $\bar{K}$ is the shared-key while $K'$ is known to all the encoder, decoder, and eavesdropper.
		It is sufficient to show that there exists a realization $(f_1,f_2,k')$ that satisfies \eqref{eqn:limsupEd<D} and \eqref{eqn:equiv>delta}.
		
		For the distortion, we exactly do as the former case with $(F_1,F_2,K')$ instead of $(F_1,F_2)$.

		For the equivocation, if we use encoding with rate $\bar{R}_0$ instead of $R_0$, according to the previous case, we have
		\begin{equation*}
			\H{X^n \mid Z^n, M_1, M_2, F_1, F_2, b}
			\geq n \H{X \mid Z, U} - n\delta.
		\end{equation*}
		Further, we obtain
		\makeatletter
		\if@twocolumn
			\begin{IEEEeqnarray*}{rCl}
				\IEEEeqnarraymulticol{3}{l}{\H{X^n \mid Z^n, M_1, M_2, F_1, F_2, K', b}} \\
				\qquad&\leq& \H{X^n \mid Z^n, M_1, M_2, F_1, F_2, b}
				+ n \bar{R}_0 - nR_0.
			\end{IEEEeqnarray*}
		\else
			\begin{equation*}
				\H{X^n \mid Z^n, M_1, M_2, F_1, F_2, K', b}
				\leq \H{X^n \mid Z^n, M_1, M_2, F_1, F_2, b}
				+ n \bar{R}_0 - nR_0.
			\end{equation*}
		\fi
		\makeatother
		Hence, we have
		\makeatletter
		\if@twocolumn
			\begin{IEEEeqnarray}{rCl}
				\IEEEeqnarraymulticol{3}{l}{\H{X^n \mid Z^n, M_1, M_2, F_1, F_2, b}} \nonumber\\
				\quad&\geq& n \left[\H{X \mid Z, U} - \I{X}{V \mid Y, U}
				+ R_0 - \delta \right] \nonumber\\
				&=& n [\I{Y}{V \mid U} - \I{Z}{V \mid U} \nonumber\\
				&&\quad + \H{X \mid Z, V} + R_0 - \delta ], \label{eqn:Delta case2, W=Y}
			\end{IEEEeqnarray}
		\else
			\begin{IEEEeqnarray}{rCl}
				\IEEEeqnarraymulticol{3}{l}{\H{X^n \mid Z^n, M_1, M_2, F_1, F_2, b}} \nonumber\\
				\qquad&\geq& n \left[\H{X \mid Z, U} - \I{X}{V \mid Y, U}
				+ R_0 - \delta \right] \nonumber\\
				&=& n [\I{Y}{V \mid U} - \I{Z}{V \mid U}
				+ \H{X \mid Z, V} + R_0 - \delta ], \label{eqn:Delta case2, W=Y} \IEEEeqnarraynumspace
			\end{IEEEeqnarray}
		\fi
		\makeatother
		where the equality follows because
		\makeatletter
		\if@twocolumn
			\begin{IEEEeqnarray*}{rCl}
				\IEEEeqnarraymulticol{3}{l}{\H{X \mid Z, U} - \I{X}{V \mid Y, U} - \H{X\mid Z,V}} \\
				\qquad&=& \I{X}{V\mid Z,U} - \I{X}{V \mid Y,U} \\
				&=& \I{X}{V\mid U} - \I{Z}{V \mid U} \\
				&&-\I{X}{V \mid U} + \I{Y}{V \mid U} \\
				&=&\I{Y}{V \mid U} - \I{Z}{V \mid U}.
			\end{IEEEeqnarray*}
		\else
			\begin{IEEEeqnarray*}{rCl}
				\IEEEeqnarraymulticol{3}{l}{\H{X \mid Z, U} - \I{X}{V \mid Y, U} - \H{X\mid Z,V}} \\
				\qquad&=& \I{X}{V\mid Z,U} - \I{X}{V \mid Y,U} \\
				&=& \I{X}{V\mid U} - \I{Z}{V \mid U}
				-\I{X}{V \mid U} + \I{Y}{V \mid U} \\
				&=&\I{Y}{V \mid U} - \I{Z}{V \mid U}.
			\end{IEEEeqnarray*}
		\fi
		\makeatother
		Therefore, there exists $(f_1, f_2, k')$ such that \eqref{eqn:limsupEd<D} and \eqref{eqn:equiv>delta} are satisfied.
	\end{itemize}

	Thus, the proof of the theorem is complete.
	We only need to prove \eqref{eqn:Final ineq}.
	
	\emph{Proof of \eqref{eqn:Final ineq}}:
	First, we remove $\tilde{R}_2$.
	Hence, from \eqref{eqn:Rf1,Rf2<2}, \eqref{eqn:Rf1,Rf2<3}, \eqref{eqn:R+Rf >2}, and \eqref{eqn:R+Rf >3}, we have
	\begin{IEEEeqnarray*}{l}
		\max\{\H{U\mid V} - R_2, \H{U, V \mid Y} - R_0 - R_1 - R_2 - \tilde{R}_1\} \nonumber\\
		\qquad<\tilde{R}_2
		<\min\{\H{U\mid X},\H{U, V \mid X} - R_0 - \tilde{R}_1\}.
	\end{IEEEeqnarray*}
	Hence, instead of \eqref{eqn:Rf1,Rf2<2}, \eqref{eqn:Rf1,Rf2<3}, \eqref{eqn:R+Rf >2}, and \eqref{eqn:R+Rf >3}, we have
	\makeatletter
	\if@twocolumn
		\begin{IEEEeqnarray}{l}
			R_2 > \H{U \mid V} - \H{U \mid X}
			= -\I{U}{V \mid X}, \label{eqn:FM:-tR2-1} \\
			R_0 + \tilde{R}_1 - R_2 
			< \H{U,V \mid X} - \H{U \mid V} \nonumber\\
			\qquad= \H{V \mid X}, \label{eqn:FM:-tR2-2} \\
			R_0 + R_1 + R_2 + \tilde{R}_1 
			> \H{U,V\mid Y} - \H{U\mid X} \nonumber\\
			\qquad= \H{V \mid Y} - \I{U}{V \mid X}, \label{eqn:FM:-tR2-3} \\
			R_1 + R_2 > \H{U,V\mid Y} - \H{U,V\mid X} \nonumber\\
			\qquad= \I{X}{U, V \mid Y}
			= \I{X}{V \mid Y}, \label{eqn:FM:-tR2-4}
		\end{IEEEeqnarray}
	\else
		\begin{IEEEeqnarray}{l}
			R_2 > \H{U \mid V} - \H{U \mid X}
			= -\I{U}{V \mid X}, \label{eqn:FM:-tR2-1} \\
			R_0 + \tilde{R}_1 - R_2 
			< \H{U,V \mid X} - \H{U \mid V}
			= \H{V \mid X}, \label{eqn:FM:-tR2-2} \\
			R_0 + R_1 + R_2 + \tilde{R}_1 
			> \H{U,V\mid Y} - \H{U\mid X}
			= \H{V \mid Y} - \I{U}{V \mid X}, \label{eqn:FM:-tR2-3} \\
			R_1 + R_2 > \H{U,V\mid Y} - \H{U,V\mid X}
			= \I{X}{U, V \mid Y}
			= \I{X}{V \mid Y}, \label{eqn:FM:-tR2-4}
		\end{IEEEeqnarray}
	\fi
	\makeatother
	where all the equalities follows from the Markov chain $U\to V\to X\to (Y,Z)$. \\
	Next, we remove $\tilde{R}_1$.
	From \eqref{eqn:Rf1,Rf2<1}, \eqref{eqn:R+Rf >1}, \eqref{eqn:R+Rf < Z indep}, \eqref{eqn:FM:-tR2-2}, and \eqref{eqn:FM:-tR2-3}, we obtain
	\makeatletter
	\if@twocolumn
		\begin{IEEEeqnarray*}{l}
			\max\{
			\H{V\mid Y,U} - R_0 - R_1, \\
			\qquad\H{V\mid Y} - \I{U}{V \mid X} - R_0 - R_1 - R_2
			\} \nonumber\\
			\qquad<\tilde{R}_1
			<\min\{
			\H{V\mid X} - R_0, \\
			\qquad\qquad\H{V\mid X,U} - R_1, 
			\H{V\mid X} - R_0 + R_2
			\}.
		\end{IEEEeqnarray*}
	\else
		\begin{IEEEeqnarray*}{l}
			\max\{
			\H{V\mid Y,U} - R_0 - R_1, 
			\H{V\mid Y} - \I{U}{V \mid X} - R_0 - R_1 - R_2
			\} \nonumber\\
			\qquad<\tilde{R}_1
			<\min\{
			\H{V\mid X} - R_0,
			\H{V\mid X,U} - R_1, 
			\H{V\mid X} - R_0 + R_2
			\}.
		\end{IEEEeqnarray*}
	\fi
	\makeatother
	Therefore, the following inequalities are substituted with \eqref{eqn:Rf1,Rf2<1}, \eqref{eqn:R+Rf >1}, \eqref{eqn:R+Rf < Z indep}, \eqref{eqn:FM:-tR2-2}, and \eqref{eqn:FM:-tR2-3}
	\makeatletter
	\if@twocolumn
		\begin{IEEEeqnarray}{rCl}
			R_1 
			&>& \H{V \mid Y, U} - \H{V \mid X} \nonumber\\
			&=& \I{X}{V \mid Y, U} - \I{U}{V \mid X}, \label{eqn:FM:-tR1-1} \\
			R_0 
			&>& \H{V \mid Y, U} - \H{V \mid X, U} \nonumber\\
			&=& \I{X}{V \mid Y, U}, \label{eqn:FM:-tR1-2} \\
			R_1 + R_2
			&>& \H{V \mid Y, U} - \H{V \mid X}, \label{eqn:FM:-tR1-3} \\
			R_1 + R_2
			&>& \H{V \mid Y} - \I{U}{V \mid X} - \H{V \mid X} \nonumber\\
			&=& \I{X}{V \mid Y} - \I{U}{V \mid X}, \label{eqn:FM:-tR1-4} \\
			R_0 + R_2
			&>& \H{V \mid Y} - \I{U}{V \mid X} - \H{V \mid X, U} \nonumber\\
			&=& \I{X}{V \mid Y}, \label{eqn:FM:-tR1-5} \\
			R_1 + 2 R_2
			&>& \H{V \mid Y} - \I{U}{V \mid X} - \H{V \mid X} \nonumber\\
			&=& \I{X}{V \mid Y} - \I{U}{V \mid X}, \label{eqn:FM:-tR1-6}
		\end{IEEEeqnarray}
	\else
		\begin{IEEEeqnarray}{rCl}
			R_1 
			&>& \H{V \mid Y, U} - \H{V \mid X}
			= \I{X}{V \mid Y, U} - \I{U}{V \mid X}, \label{eqn:FM:-tR1-1} \\
			R_0 
			&>& \H{V \mid Y, U} - \H{V \mid X, U}
			= \I{X}{V \mid Y, U}, \label{eqn:FM:-tR1-2} \\
			R_1 + R_2
			&>& \H{V \mid Y, U} - \H{V \mid X}, \label{eqn:FM:-tR1-3} \\
			R_1 + R_2
			&>& \H{V \mid Y} - \I{U}{V \mid X} - \H{V \mid X}
			= \I{X}{V \mid Y} - \I{U}{V \mid X}, \label{eqn:FM:-tR1-4} \\
			R_0 + R_2
			&>& \H{V \mid Y} - \I{U}{V \mid X} - \H{V \mid X, U}
			= \I{X}{V \mid Y}, \label{eqn:FM:-tR1-5} \\
			R_1 + 2 R_2
			&>& \H{V \mid Y} - \I{U}{V \mid X} - \H{V \mid X}
			= \I{X}{V \mid Y} - \I{U}{V \mid X}, \label{eqn:FM:-tR1-6}
		\end{IEEEeqnarray}
	\fi
	\makeatother
	where all the equalities follows from the Markov chain $U\to V\to X\to (Y,Z)$.
	Thus, all the inequalities are \eqref{eqn:FM:-tR2-1}, \eqref{eqn:FM:-tR2-4}, \eqref{eqn:FM:-tR1-1}, \eqref{eqn:FM:-tR1-2}, \eqref{eqn:FM:-tR1-3}, \eqref{eqn:FM:-tR1-4}, \eqref{eqn:FM:-tR1-5}, and \eqref{eqn:FM:-tR1-6}, where \eqref{eqn:FM:-tR2-1}, \eqref{eqn:FM:-tR1-3}, \eqref{eqn:FM:-tR1-4}, and \eqref{eqn:FM:-tR1-6} are redundant because of $R_2 \geq 0$, \eqref{eqn:FM:-tR1-1}, \eqref{eqn:FM:-tR2-4}, and \eqref{eqn:FM:-tR1-4}, respectively.
	Finally, by selecting $R_1 = R - R_2$, from \eqref{eqn:FM:-tR2-4}, \eqref{eqn:FM:-tR1-1}, and \eqref{eqn:FM:-tR1-5} we obtain,
	\begin{IEEEeqnarray}{l}
		R > \I{X}{V \mid Y}, \label{FM:-R2-1}\\
		R - R_2 > \I{X}{V \mid Y, U} - \I{U}{V \mid X}, \label{FM:-R2-2}\\
		R_0 + R_2 > \I{X}{V \mid Y}. \label{FM:-R2-3}
	\end{IEEEeqnarray}
	Next, we remove $R_2$.
	Hence, instead of \eqref{FM:-R2-2} and \eqref{FM:-R2-3}, we have
	\makeatletter
	\if@twocolumn
		\begin{IEEEeqnarray*}{l}
			\I{X}{V \mid Y} - R_0
			< R_2 \\
			\qquad < R - \I{X}{V \mid Y, U} + \I{U}{V \mid X} \\
			\Rightarrow R + R_0 > \I{X}{V \mid Y} + \I{X}{V \mid Y, U} - \I{U}{V \mid X},
		\end{IEEEeqnarray*}
	\else
		\begin{IEEEeqnarray*}{l}
			\I{X}{V \mid Y} - R_0
			< R_2
			< R - \I{X}{V \mid Y, U} + \I{U}{V \mid X} \\
			\qquad\Rightarrow R + R_0 > \I{X}{V \mid Y} + \I{X}{V \mid Y, U} - \I{U}{V \mid X},
		\end{IEEEeqnarray*}
	\fi
	\makeatother
	which is redundant due to \eqref{eqn:FM:-tR1-2} and \eqref{FM:-R2-1}.
	Thus, \eqref{eqn:Final ineq} follows.

\subsection{Converse}
	To prove the converse part, we assume that there exists a sequence of source codes with rate $R$ such that \eqref{eqn:limsupEd<D} and \eqref{eqn:equiv>delta} are satisfied, and we will show that we can identify auxillary $\rv$s assignment, with conditional distribution $p(v | x) \: p(u | v) \: p(\hat{x} | v, y)$, that satisfies \eqref{eqn:R(SI+SR)}, \eqref{eqn:Delta(SI+SR)}, and \eqref{eqn:D(SI+SR)}.
	For $i=1,\ldots,n$, we assign 
	\begin{equation} \label{eqn:SI+SR:auxRVs}
		\begin{cases}
		U_i := \left(Y_{i+1}^n, Z^{i-1}, M\right), \\
		V_i := \left(X^{i-1}, Y^{i-1}, Y_{i+1}^n, Z^{i-1}, M, K \right), \\
		Q\sim\mathrm{Unif}\{1,\ldots,n\}, \\
		V := (V_Q, Q), \\
		Q := (U_Q, Q),
		\end{cases}
	\end{equation}
	where $Q$ is independent of $(M, K, X^n, Y^n, Z^n, \hat{X}^n)$.
	It can be seen that we have the Markov chain $U_i \rightarrow V_i \rightarrow X_i \rightarrow (Y_i, Z_i)$, for $i=1,\ldots,n$ (see Fig. \ref{fig:converse}).
	
	\begin{figure}
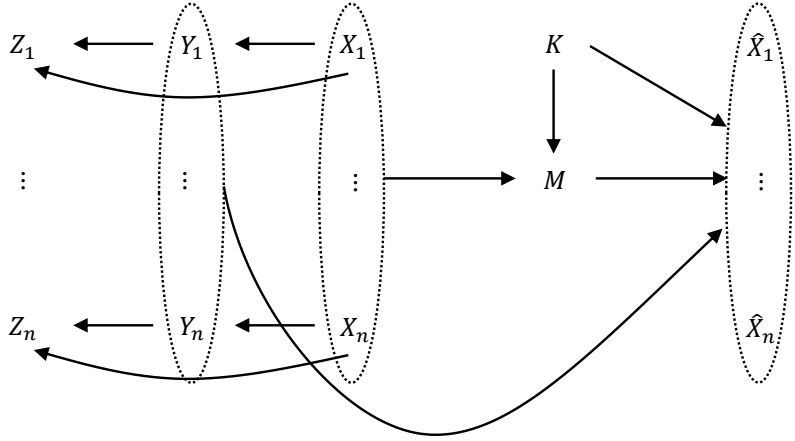

		\centering
		\makeatletter
		\if@twocolumn
			\includegraphics[trim = 0 20 0 30, clip, width=\columnwidth]{converse1}
		\else
			\includegraphics[trim = 0 20 0 30, clip, width=0.7\columnwidth]{converse1}
		\fi
		\makeatother
		\caption{The graphical representation of the problem.}
		\label{fig:converse}
	\end{figure}

	\emph{Proof of \eqref{eqn:R(SI+SR)}}:
	Because the coding rate is $R$, the following series of inequalities hold:
	\makeatletter
	\if@twocolumn
		\begin{IEEEeqnarray}{rCl}
			n R 
			&\geq& \H{M}
			\geq \H{M \mid Y^n, Z^{i-1}, K} \nonumber\\
			&\geq& \I{X^n}{M \mid Y^n, Z^{i-1}, K} \nonumber\\
			&=& \sum_{i=1}^n{\I{X_i}{M \mid X^{i-1}, Y^n, Z^{i-1}, K}} \nonumber\\
			&\stackrel{(a)}{=}& \sum_{i=1}^n{\I{X_i}{X^{i-1}, Y^{i-1}, Y_{i+1}^n, Z^{i-1}, M, K \mid Y_i}} \nonumber\\
			&=& \sum_{i=1}^n{\I{X_i}{V_i \mid Y_i}} \nonumber\\
			&\stackrel{(b)}{=}& n \I{X_Q}{V_Q \mid Y_Q, Q} \nonumber\\
			&\stackrel{(c)}{=}& n \I{X_Q}{V_Q, Q \mid Y_Q} \nonumber\\
			&\stackrel{(d)}{=}& n \I{X}{V \mid Y}, \label{eqn:SI+SR:R>I(X;V|Y)}
		\end{IEEEeqnarray}
	\else
		\begin{IEEEeqnarray}{rCl}
			n R 
			&\geq& \H{M}
			\geq \H{M \mid Y^n, Z^{i-1}, K}
			\geq \I{X^n}{M \mid Y^n, Z^{i-1}, K} \nonumber\\
			&=& \sum_{i=1}^n{\I{X_i}{M \mid X^{i-1}, Y^n, Z^{i-1}, K}} \nonumber\\
			&\stackrel{(a)}{=}& \sum_{i=1}^n{\I{X_i}{X^{i-1}, Y^{i-1}, Y_{i+1}^n, Z^{i-1}, M, K \mid Y_i}} \nonumber\\
			&=& \sum_{i=1}^n{\I{X_i}{V_i \mid Y_i}} \nonumber\\
			&\stackrel{(b)}{=}& n \I{X_Q}{V_Q \mid Y_Q, Q} \nonumber\\
			&\stackrel{(c)}{=}& n \I{X_Q}{V_Q, Q \mid Y_Q} \nonumber\\
			&\stackrel{(d)}{=}& n \I{X}{V \mid Y}, \label{eqn:SI+SR:R>I(X;V|Y)}
		\end{IEEEeqnarray}
	\fi
	\makeatother
	\sloppy where $(a)$ follows from $(X^n, Y^n, Z^n)$ being i.i.d. and $K \indep (X^n, Y^n, Z^n)$, as a result, $(X_i, Y_i) \indep (X^{i-1}, Y^{i-1}, Y_{i+1}^n, Z^{i-1}, K)$;
	$(b)$ follows from $Q \indep (X^n, Y^n)$ and $Q$ being uniformly distributed;
	$(c)$ follows from independency of $Q$ and $(X_Q, Y_Q)$ which is the result of $(X^n, Y^n)$ being i.i.d., $Q \indep (X^n, Y^n)$ and uniformly distributed;
	and $(d)$ follows from the Markov chain $V \rightarrow (X_Q, Y_Q) \rightarrow Z_Q$ and $(X_Q, Y_Q, Z_Q) \sim p(x,y,z)$, so, we can substitute $X_Q$, $Y_Q$, and $Z_Q$ with $X$, $Y$, and $Z$, respectively.
	
	\emph{Proof of \eqref{eqn:Delta(SI+SR)}}:
	From \eqref{eqn:equiv>delta}, we obtain that, for any given small $\epsilon$, if $n$ is large enough, we have
	\makeatletter
	\if@twocolumn
		\begin{IEEEeqnarray}{rCl}
			n(\Delta - \epsilon)
			&\leq& \H{X^n \mid M, Z^n} \nonumber\\
			&=& \H{X^n \mid Z^n} - \I{X^n}{M \mid Z^n} \nonumber\\
			&\stackrel{(a)}{=}& \H{X^n \mid Z^n} - \I{X^n}{M} + \I{Z^n}{M} \nonumber\\
			&\stackrel{(b)}{=}& \H{X^n \mid Z^n} \nonumber\\
			&&- \I{X^n}{M \mid K} + \I{X^n}{K \mid M} + \I{Z^n}{M} \nonumber\\
			&& - \I{Y^n}{M} + \I{Y^n}{M \mid K} - \I{Y^n}{K \mid M} \nonumber\\
			&=& \H{X^n \mid Z^n} \nonumber\\
			&& - \left[\I{X^n}{M \mid K} - \I{Y^n}{M \mid K} \right] \nonumber\\
			&& + \I{Z^n}{M} - \I{Y^n}{M} \nonumber\\
			&& + \I{X^n}{K \mid M} - \I{Y^n}{K \mid M}, \label{eqn:SI+SR:Delta UB multiletter}
		\end{IEEEeqnarray}
	\else
		\begin{IEEEeqnarray}{rCl}
			n(\Delta - \epsilon)
			&\leq& \H{X^n \mid M, Z^n}
			= \H{X^n \mid Z^n} - \I{X^n}{M \mid Z^n} \nonumber\\
			&\stackrel{(a)}{=}& \H{X^n \mid Z^n} - \I{X^n}{M} + \I{Z^n}{M} \nonumber\\
			&\stackrel{(b)}{=}& \H{X^n \mid Z^n} \nonumber\\
			&&- \I{X^n}{M \mid K} + \I{X^n}{K \mid M} + \I{Z^n}{M} \nonumber\\
			&& - \I{Y^n}{M} + \I{Y^n}{M \mid K} - \I{Y^n}{K \mid M} \nonumber\\
			&=& \H{X^n \mid Z^n} \nonumber\\
			&& - \left[\I{X^n}{M \mid K} - \I{Y^n}{M \mid K} \right] \nonumber\\
			&& + \I{Z^n}{M} - \I{Y^n}{M} \nonumber\\
			&& + \I{X^n}{K \mid M} - \I{Y^n}{K \mid M}, \label{eqn:SI+SR:Delta UB multiletter}
		\end{IEEEeqnarray}
	\fi
	\makeatother
	where $(a)$ follows from the Markov chain $M \to X^n \to Z^n$; and
	$(b)$ follows from $X^n \indep K$ and $Y^n \indep K$.
	Next, we bound each term in \eqref{eqn:SI+SR:Delta UB multiletter}, as follows:
	\makeatletter
	\if@twocolumn
		\begin{IEEEeqnarray}{rCl}
			\IEEEeqnarraymulticol{3}{l}{\I{X^n}{M \mid K} - \I{Y^n}{M \mid K}} \nonumber\\
			\quad&\stackrel{(a)}{=}& \I{X^n}{M \mid Y^n, K} \nonumber\\
			&=& \sum_{i=1}^n{\I{X_i}{M \mid X^{i-1}, Y^n, K}} \nonumber\\
			&\stackrel{(b)}{=}& \sum_{i=1}^n{\I{X_i}{M \mid X^{i-1}, Y^n, Z^{i-1}, K}} \nonumber\\
			&\stackrel{(c)}{=}& \sum_{i=1}^n{\I{X_i}{X^{i-1}, Y^{i-1}, Y_{i+1}^n, Z^{i-1}, M, K \mid Y_i}} \nonumber\\
			&=& \sum_{i=1}^n{\I{X_i}{V_i \mid Y_i}}, \label{eqn:SI+SR:multi->single eq1}
		\end{IEEEeqnarray}
	\else
		\begin{IEEEeqnarray}{rCl}
			\I{X^n}{M \mid K} - \I{Y^n}{M \mid K}
			&\stackrel{(a)}{=}& \I{X^n}{M \mid Y^n, K} \nonumber\\
			&=& \sum_{i=1}^n{\I{X_i}{M \mid X^{i-1}, Y^n, K}} \nonumber\\
			&\stackrel{(b)}{=}& \sum_{i=1}^n{\I{X_i}{M \mid X^{i-1}, Y^n, Z^{i-1}, K}} \nonumber\\
			&\stackrel{(c)}{=}& \sum_{i=1}^n{\I{X_i}{X^{i-1}, Y^{i-1}, Y_{i+1}^n, Z^{i-1}, M, K \mid Y_i}} \nonumber\\
			&=& \sum_{i=1}^n{\I{X_i}{V_i \mid Y_i}}, \label{eqn:SI+SR:multi->single eq1}
		\end{IEEEeqnarray}
	\fi
	\makeatother
	where $(a)$ follows from $M \to (X^n, K) \to Y^n$;
	whereas $(b)$ and $(c)$ follow from $(X^n, Y^n, Z^n)$ being i.i.d., and $M \to (X^n, K) \to (Y^n, Z^n)$, as a result, $(X_i, Y_{i+1}^n, M) \to (X^{i-1}, K) \to (Y^{i-1}, Z^{i-1})$. \\
	For the next term in \eqref{eqn:SI+SR:Delta UB multiletter}, we have
	\makeatletter
	\if@twocolumn
		\begin{IEEEeqnarray}{rCl}
			\IEEEeqnarraymulticol{3}{l}{\I{Z^n}{M} - \I{Y^n}{M}} \nonumber\\
			\quad&=& \sum_{i=1}^n{\I{Z_i}{M \mid Z^{i-1}} - \I{Y_i}{M \mid Y_{i+1}^n}} \nonumber\\
			&\stackrel{(a)}{=}& \sum_{i=1}^n{\I{Z_i}{Y_{i+1}^n, M \mid Z^{i-1}} - \I{Y_i}{Z^{i-1}, M \mid Y_{i+1}^n}} \nonumber\\
			&\stackrel{(b)}{=}& \sum_{i=1}^n{\I{Z_i}{Y_{i+1}^n, Z^{i-1}, M} - \I{Y_i}{Y_{i+1}^n, Z^{i-1}, M}} \nonumber\\
			&=& \sum_{i=1}^n{\I{Z_i}{U_i} - \I{Y_i}{U_i}}, \label{eqn:SI+SR:multi->single eq2}
		\end{IEEEeqnarray}
	\else
		\begin{IEEEeqnarray}{rCl}
			\I{Z^n}{M} - \I{Y^n}{M}
			&=& \sum_{i=1}^n{\I{Z_i}{M \mid Z^{i-1}} - \I{Y_i}{M \mid Y_{i+1}^n}} \nonumber\\
			&\stackrel{(a)}{=}& \sum_{i=1}^n{\I{Z_i}{Y_{i+1}^n, M \mid Z^{i-1}} - \I{Y_i}{Z^{i-1}, M \mid Y_{i+1}^n}} \nonumber\\
			&\stackrel{(b)}{=}& \sum_{i=1}^n{\I{Z_i}{Y_{i+1}^n, Z^{i-1}, M} - \I{Y_i}{Y_{i+1}^n, Z^{i-1}, M}} \nonumber\\
			&=& \sum_{i=1}^n{\I{Z_i}{U_i} - \I{Y_i}{U_i}}, \label{eqn:SI+SR:multi->single eq2}
		\end{IEEEeqnarray}
	\fi
	\makeatother
	where $(a)$ follows from Csisz\'ar sum identity \cite[p. 25]{ElGamal11};
	and $(b)$ follows from $(X^n, Y^n, Z^n)$ being i.i.d., as a result, $(Y_i, Z_i) \indep (Y_{i+1}^n, Z^{i-1})$, for all $i=1,\ldots,n$. \\
	For the next term in \eqref{eqn:SI+SR:Delta UB multiletter}, we obtain
	\makeatletter
	\if@twocolumn
		\begin{IEEEeqnarray}{rCl}
			\I{X^n}{K \mid M} - \I{Y^n}{K \mid M}
			&\stackrel{(a)}{=}& \I{X^n}{K \mid Y^n, M} \nonumber\\ 
			&\leq& \H{K}
			= n R_0, \label{eqn:SI+SR:multi->single ineq1}
		\end{IEEEeqnarray}
	\else
		\begin{equation} \label{eqn:SI+SR:multi->single ineq1}
			\I{X^n}{K \mid M} - \I{Y^n}{K \mid M}
			\stackrel{(a)}{=} \I{X^n}{K \mid Y^n, M} 
			\leq \H{K}
			= n R_0,
		\end{equation}
	\fi
	\makeatother
	where $(a)$ follows from $(X^n, Y^n) \indep K$ and the Markov chain $Y^n \to (X^n, K) \to M$.
	Utilizing this result, we further obtain
	\makeatletter
	\if@twocolumn
		\begin{IEEEeqnarray}{rCl}
			\IEEEeqnarraymulticol{3}{l}{\I{X^n}{K \mid M} - \I{Y^n}{K \mid M}} \nonumber\\
			\quad&=& \I{X^n}{K \mid Y^n, M} \nonumber\\
			&=& \sum_{i=1}^n{\I{X_i}{K \mid X^{i-1}, Y^n, M}} \nonumber\\
			&\stackrel{(a)}{=}& \sum_{i=1}^n{\I{X_i}{K \mid X^{i-1}, Y^n, Z^{i-1}, M}} \nonumber\\
			&\leq& \sum_{i=1}^n{\I{X_i}{X^{i-1}, Y^{i-1}, K \mid Y_i, Y_{i+1}^n, Z^{i-1}, M}} \nonumber\\
			&=& \sum_{i=1}^n{\I{X_i}{V_i \mid Y_i, U_i}}, \label{eqn:SI+SR:multi->single ineq2}
		\end{IEEEeqnarray}
	\else
		\begin{IEEEeqnarray}{rCl}
			\I{X^n}{K \mid M} - \I{Y^n}{K \mid M}
			&=& \I{X^n}{K \mid Y^n, M} \nonumber\\
			&=& \sum_{i=1}^n{\I{X_i}{K \mid X^{i-1}, Y^n, M}} \nonumber\\
			&\stackrel{(a)}{=}& \sum_{i=1}^n{\I{X_i}{K \mid X^{i-1}, Y^n, Z^{i-1}, M}} \nonumber\\
			&\leq& \sum_{i=1}^n{\I{X_i}{X^{i-1}, Y^{i-1}, K \mid Y_i, Y_{i+1}^n, Z^{i-1}, M}} \nonumber\\
			&=& \sum_{i=1}^n{\I{X_i}{V_i \mid Y_i, U_i}}, \label{eqn:SI+SR:multi->single ineq2}
		\end{IEEEeqnarray}
	\fi
	\makeatother
	where $(a)$ follows from $(X^n, Y^n, Z^n)$ being i.i.d. and $K \to (X^n, M) \to (Y^n, Z^n)$, which is obtained from $M \to (X^n, K) \to (Y^n, Z^n)$ and $K \indep (X^n, Y^n, Z^n)$, as a result, $(X_i, Y_{i+1}^n, K) \to (X^{i-1}, M) \to (Y^{i-1}, Z^{i-1})$ (see Fig. \ref{fig:converse} and marginalize over $\hat{X}^n$).\\
	Therefore, from \eqref{eqn:SI+SR:Delta UB multiletter}, \eqref{eqn:SI+SR:multi->single eq1}, \eqref{eqn:SI+SR:multi->single eq2}, \eqref{eqn:SI+SR:multi->single ineq1}, \eqref{eqn:SI+SR:multi->single ineq2}, and the fact that $(X^n, Z^n)$ is i.i.d., we obtain that
	\begin{IEEEeqnarray}{rCl}
		n(\Delta - \epsilon)
		&\leq& \sum_{i=1}^n
		\H{X_i \mid Z_i} 
		- \I{X_i}{V_i \mid Y_i} \nonumber\\
		&&\qquad+ \I{Z_i}{U_i} - \I{Y_i}{U_i} \nonumber\\
		&&\qquad+ \min\{R_0, \I{X_i}{V_i \mid Y_i, U_i}\} \nonumber\\ 
		&=& n[\H{X \mid Z} 
		- \I{X_Q}{V_Q \mid Y_Q, Q} \nonumber\\
		&&\quad+ \I{Z_Q}{U_Q \mid Q} - \I{Y_Q}{U_Q \mid Q} \nonumber\\
		&&\quad+ \min\{R_0, \I{X_Q}{V_Q \mid Y_Q, U_Q, Q}\}] \nonumber\\
		&=& n[\H{X \mid Z} 
		- \I{X_Q}{V_Q, Q \mid Y_Q} \nonumber\\
		&&\quad+ \I{Z_Q}{U_Q, Q} - \I{Y_Q}{U_Q, Q} \nonumber\\
		&&\quad+ \min\{R_0, \I{X_Q}{V_Q, \mid Y_Q, U_Q, Q}\}] \nonumber\\
		&=& n[ \H{X \mid Z} 
		- \I{X}{V \mid Y} \nonumber\\
		&&\quad + \I{Z}{U} - \I{Y}{U} \nonumber\\
		&&\quad +\min\{R_0, \I{X}{V \mid Y, U}\}], \IEEEeqnarraynumspace\label{eqn:Delta UB final 1}
	\end{IEEEeqnarray}
	where the last three steps follow the same arguments of the last three steps of \eqref{eqn:SI+SR:R>I(X;V|Y)}.
	Using \eqref{eqn:SI+SR:Delta UP simple1} and \eqref{eqn:SI+SR:Delta UP simple2} from Lemma \ref{lmm:=}, \eqref{eqn:Delta(SI+SR)} follows from \eqref{eqn:Delta UB final 1}.

	\emph{Proof of \eqref{eqn:D(SI+SR)}}:
	From \eqref{eqn:limsupEd<D}, we obtain that, for any given small $\epsilon$, if $n$ is large enough, we have
	\begin{IEEEeqnarray*}{rCl}
		D + \epsilon
		&\geq& \E{d(X^n,\hat{X}^n)}
		= \frac{1}{n} \sum_{i=1}^n{\E{d(X_i, \hat{X}_i)}} \\
		&\stackrel{(a)}{=}& \E{d(X_Q, \hat{X}_Q)} \\
		&\stackrel{(b)}{=}& \E{d(X, \hat{X})},
	\end{IEEEeqnarray*}
	where $(a)$ follows from $Q \indep (X^n, \hat{X}^n)$;
	for $(b)$ recall the Markov chain $(X^n, Z^n) \rightarrow (Y^n, M, K) \rightarrow \hat{X}^n$, as a result, we have $(X_i, Z_i) \rightarrow (V_i, Y_i) \rightarrow \hat{X}_i$, so, $(X_Q, Z_Q) \rightarrow (V, Y_Q) \rightarrow \hat{X}_Q$; thus, it follows by renaming $\hat{X}_Q$ with $\hat{X}$ and utilizing the fact that $X_Q \sim p(x)$.
	
	To show that it is sufficient for $\hat{X}$ to be a function of $(V,Y)$, we do the following
	\begin{IEEEeqnarray*}{rCl}
		\E{d(X,\hat{X})}
		&=& \E{\E{d(X,\hat{X}) \mid Y,V}} \\
		&\geq& \E{\min_{\hat{x}} \E{d(X, \hat{x}(Y,V)) \mid Y,V}}.
	\end{IEEEeqnarray*}
	Hence, the result follows.
	
\subsection{Cardinality bounds}
	To show the cardinality bounds, we use the method explained in \cite[Appendix C]{ElGamal11}.
	First, we assume that $p(x|v)$ is fixed.
	Without loss of generality, we assume that $\mathcal{X}=\{1,\ldots,|\mathcal{X}|\}$.
	Consider the following functions over the set of all $\pmf$s $p(v|u)$ on $\mathcal{V}$:
	\begin{equation*}
		\begin{cases}
			p(x|u) \qquad x=1,\ldots,|\mathcal{X}|-1, \\
			\H{X \mid Y, V, U=u}, \\
			\I{Y}{V \mid U=u}, \\
			\I{Z}{V \mid U=u}, \\
			\H{X \mid Z, V, U=u}, \\
			\H{X\mid Z,U=u}.
		\end{cases}
	\end{equation*}
	Therefore, from the support lemma \cite[Appendix C]{ElGamal11}, we obtain that for any $U\sim F(u)$, there exists $U'$ with $\pmf$ of cardinlity at most $|\mathcal{X}|+4$ such that the following functions are preserved:
	\begin{equation*}
		\begin{cases}
			p(x) \quad x\in\mathcal{X} \\
			\quad\Rightarrow p(x,y,z) = p(x)p(y,z|x) \quad (x,y,z)\in\mathcal{X}\times\mathcal{Y}\times\mathcal{Z}, \\
			\H{X \mid Y, V, U} = \H{X \mid Y, V} \\
			\quad\Rightarrow \I{X}{V \mid Y} = \H{X\mid Y} - \H{X \mid Y, V}, \\
			\I{Y}{V \mid U}, \\
			\I{Z}{V \mid U}, \\
			\H{X \mid Z, V, U} = \H{X\mid Z, V}, \\
			\H{X\mid Z,U}.
		\end{cases}
	\end{equation*}
	
	Let $V'$ denote the corresponding $\rv$ after choosing $U'$. 
	For bounding the cardinality of $V$, for each $u'\in\mathcal{U}'$, consider the following functions over the set of all $\pmf$s $p(x|v',x')$:
	\begin{equation} \label{eqn:CB2}
		\begin{cases}
			p(x|u',v') \qquad x=1,\ldots,|\mathcal{X}|-1, \\
			\H{X \mid Y, V'=v', U'=u'}, \\
			\H{Y \mid V'=v', U'=u'}, \\
			\H{Z \mid V'=v', U'=u'}, \\
			\H{X \mid Z, V'=v', U'=u'}.
		\end{cases}
	\end{equation}
	Therefore, from the support lemma \cite[Appendix C]{ElGamal11}, we obtain that for any $u'\in\mathcal{U}'$ and $V'\sim F(v'|u')$, there exists $V''$ with $\pmf$ of cardinlity at most $|\mathcal{X}|+3$ such that the following functions are preserved:
	\makeatletter
	\if@twocolumn
		\begin{equation*}
			\begin{cases}
				p(x|u') \quad x\in\mathcal{X} \\
				\Rightarrow p(x,y,z|u') = p(x|u')p(y,z|x) \quad (x,y,z)\in\mathcal{X}\times\mathcal{Y}\times\mathcal{Z}, \\
				\H{X \mid Y, V, U'=u'}, \\
				\H{Y \mid V', U'=u'} \\
				\quad\Rightarrow\H{Y|U'=u'} - \H{Y \mid V', U'=u'}, \\
				\H{Z \mid V', U'=u'} \\
				\quad\Rightarrow\H{Z\mid U'=u'} - \H{Z \mid V', U'=u'}, \\
				\H{X \mid Z, V, U'=u'}.
			\end{cases}
		\end{equation*}
	\else
		\begin{equation*}
			\begin{cases}
				p(x|u') \quad x\in\mathcal{X} \\
				\quad\Rightarrow p(x,y,z|u') = p(x|u')p(y,z|x) \quad (x,y,z)\in\mathcal{X}\times\mathcal{Y}\times\mathcal{Z}, \\
				\H{X \mid Y, V, U'=u'}, \\
				\H{Y \mid V', U'=u'} \\
				\quad\Rightarrow\H{Y|U'=u'} - \H{Y \mid V', U'=u'}, \\
				\H{Z \mid V', U'=u'} \\
				\quad\Rightarrow\H{Z\mid U'=u'} - \H{Z \mid V', U'=u'}, \\
				\H{X \mid Z, V, U'=u'}.
			\end{cases}
		\end{equation*}
	\fi
	\makeatother
	However, to have the Markov chain $U\to V\to X\to (Y,Z)$, we consider $V'''=(V'', U')$ instead of $V''$.
	In this case, the cardinality of $V'''$ becomes $(|\mathcal{X}|+4)(|\mathcal{X}|+3)$.
	With this change of variable, all the variables in \eqref{eqn:CB2} remains unaffected.
	Thus, the cardinality bound is proved.

\makeatletter
\if@twocolumn
\else
\section{Proof of Remark \ref{rmk:Ri<R<Ro}} \label{sec:prf:rmk}
\emph{Proof of $\mathcal{R}^* \subseteq \mathcal{R}_\mathrm{out}$}:
	From \cite[Propositions 1]{Chia13}, we obtain $\mathcal{R}_\mathrm{out}$ contains the tuples $(R,R_0,D,\Delta)$ such that
	\begin{IEEEeqnarray}{l}
		R \geq \I{X}{U',V' \mid K'}, \label{eqn:O:R(SI+SR)}\\
		\Delta \leq \min\{\H{X \mid Z}, 
		\H{X \mid Z, V', U'} + \I{K'}{V' \mid U'} \nonumber\\
		\qquad\qquad - \I{Z}{V' \mid U'} + \H{K' \mid U',V',X,Z} \}, \label{eqn:O:Delta(SI+SR)}\\
		D \geq \E{d(X, \hat{X}')} \label{eqn:O:D(SI+SR)},
	\end{IEEEeqnarray}
	for some conditional $\pmf$ $p(u',v' | x,k')$ and a function $\hat{x}'(u',v',k')$.
	Let $(U,\hat{X})$ satisfy conditions in Corollary \ref{cor:Y=0}.
	By selecting $U'=U$, $V'=(U,\hat{X})$, and $\hat{X}' = \hat{X}$, all of them independent of $K'$, \eqref{eqn:O:R(SI+SR)}, \eqref{eqn:O:Delta(SI+SR)}, and \eqref{eqn:O:D(SI+SR)} follow from \eqref{eqn:Y=0:R(SI+SR)}, \eqref{eqn:Y=0:Delta(SI+SR)}, and \eqref{eqn:Y=0:D(SI+SR)}, respectively, utilizing the facts that $\H{K'}=R_0$ and $\H{X\mid Z, U} \leq \H{X\mid Z}$.
	
\emph{Proof of $\mathcal{R}_\mathrm{in} \subseteq \mathcal{R}^*$}:
	From \cite[Propositions 2]{Chia13}, we obtain $\mathcal{R}_\mathrm{in}$ contains the tuples $(R,R_0,D,\Delta)$ such that
	\begin{IEEEeqnarray}{l}
		R > \I{X}{U',V' \mid K'}, \label{eqn:I:R(SI+SR)}\\
		\Delta < \min\{\H{X \mid Z, U'}, 
		\H{X \mid Z, U'} - \I{X}{V' \mid U', K'} \nonumber\\
		\qquad\qquad + \H{K' \mid U',V',X,Z} \}, \label{eqn:I:Delta(SI+SR)}\\
		D > \E{d(X, \hat{X}')}, \label{eqn:I:D(SI+SR)}
	\end{IEEEeqnarray}
	for some conditional $\pmf$ $p(u',v' | x,k')$ and a function $\hat{x}'(u',v',k')$.
	Let $(U',V',\hat{X}')$ satisfies the above inequalities of $\mathcal{R}_\mathrm{in}$.
	We select $U=U'$, $V=(U',V',K')$, and $\hat{x}(v) = \hat{x}'(u',v',k')$.
	Note that the Markov chain $U \to V \to X \to Z$ is satisfied due to $K\indep (X,Z)$.
	Hence, we obtain
	\begin{IEEEeqnarray*}{rCl}
		R &>& \I{X}{V \mid K'} \stackrel{(a)}{\geq} \I{X}{V} \\
		\Delta &<& \H{X \mid Z, U}, \\
		\Delta &<& \H{X \mid Z', U'} - \I{X}{V' \mid U', K'} \nonumber\\
		&& + \H{K' \mid U',V',X,Z} \\
		&=& \H{X \mid Z', U'} - \I{X}{V', K' \mid U'} \nonumber\\
		&& + \I{X}{K' \mid U'} + \H{K' \mid U',V',X,Z} \\
		&\leq& \H{X \mid Z', U'} - \I{X}{V', K' \mid U'} \nonumber\\
		&& + \I{X}{K' \mid U'} + \H{K' \mid U',X} \\
		&=& \H{X \mid Z', U'} - \I{X}{V', K' \mid U'}
		+ \H{K' \mid U'} \\
		&\stackrel{(b)}{\leq}& \H{X \mid Z', U'} - \I{X}{V', K' \mid U'} + R_0 \\
		&=& \H{X \mid Z, U} - \I{X}{V \mid U} + R_0 \\
		D &>& \E{d(X, \hat{X}')}
		= \E{d(X, \hat{X})}, \label{eqn:I:D(SI+SR)}
	\end{IEEEeqnarray*}
	where $(a)$ follows from $K' \indep X$;
	and $(b)$ follows from $\H{K'} = R_0$.
	Therefore, we obtain \eqref{eqn:R(SI+SR)}, \eqref{eqn:Delta(SI+SR)}, and \eqref{eqn:D(SI+SR)} for $Y=\emptyset$ utilizing \eqref{eqn:SI+SR:Delta UP simple1} in Lemma \ref{lmm:=}.
	It is the same as $\mathcal{R}^*$ as it is shown in Corollary \ref{cor:Y=0}.
\fi
\makeatother

\section{Proof of Lemma \ref{lmm:=}} \label{prf:lmm:=}
\emph{Proof of \eqref{eqn:SI+SR:Delta UP simple2}}:
	\begin{IEEEeqnarray*}{rCl}
		\IEEEeqnarraymulticol{3}{l}{\H{X \mid Z}
		- \I{X}{V \mid Y} 
		+ \I{Z}{U} - \I{Y}{U}} \nonumber\\
		\quad &=& \H{X \mid Z, V} 
		+ \I{X}{V \mid Z} \nonumber\\
		&&- \I{X}{V \mid Y}
		+ \I{Z}{U} - \I{Y}{U} \nonumber\\
		&\stackrel{(a)}{=}& \H{X \mid Z, V}
		+ \I{X}{V} - \I{Z}{V} \nonumber\\
		&& - \I{X}{V} + \I{Y}{V}
		+ \I{Z}{U} - \I{Y}{U} \nonumber\\
		&\stackrel{(b)}{=}& \H{X \mid Z, V}
		+ \I{Y}{V \mid U}
		- \I{Z}{V \mid U},
	\end{IEEEeqnarray*}
	where $(a)$ follows from $V \to X \to Z$ and $V \to X \to Y$;
	and $(b)$ follows from $U \to V \to Y$ and $U \to V \to Z$.
	
\emph{Proof of \eqref{eqn:SI+SR:Delta UP simple1}}:
	\begin{IEEEeqnarray*}{rCl}
		\IEEEeqnarraymulticol{3}{l}{\H{X \mid Z}
		- \I{X}{V \mid Y}} \nonumber\\
		&&+ \I{Z}{U} - \I{Y}{U}
		+ \I{X}{V \mid Y, U} \nonumber\\
		\quad &\stackrel{(a)}{=}& \H{X \mid Z} 
		- \I{X}{U \mid Y}
		+ \I{Z}{U} - \I{Y}{U} \nonumber\\
		&\stackrel{(b)}{=}& \H{X \mid Z}
		- \I{X}{U}
		+ \I{Z}{U} \nonumber\\
		&\stackrel{(c)}{=}& \H{X \mid Z}
		- \I{X}{U \mid Z} \nonumber\\
		&=& \H{X \mid U, Z},
	\end{IEEEeqnarray*}
	where $(a)$ follows from the Markov chain $U \to V \to (X, Y)$;
	$(b)$ follows from the Markov chain $U \to X \to Y$;
	and $(c)$ follows from the Markov chain $U \to X \to Z$.
	Finally, \eqref{eqn:SI+SR:Delta UP simple1} follows from \eqref{eqn:SI+SR:Delta UP simple2}.

\emph{Proof of \eqref{eqn:VPForm}}:
	\begin{IEEEeqnarray*}{rCl}
		\IEEEeqnarraymulticol{3}{l}{\H{X \mid Z, U} 
		- \I{X}{V \mid Y, U}} \\
		\quad &\stackrel{(a)}{=}& \H{X \mid Y, V}
		+ \H{X \mid Z,U} 
		- \H{X \mid Y,U} \\
		&=& \H{X \mid Y, V}
		- \I{X}{Z \mid U}
		+ \I{X}{Y \mid U}.
	\end{IEEEeqnarray*}
	where $(a)$ follows from the Markov chain $U \to V \to (X,Y)$.
Finally, \eqref{eqn:VPForm} follows from \eqref{eqn:SI+SR:Delta UP simple1}.
\fi
\makeatother
\end{document}